\documentclass{aa}

\usepackage{graphicx}
\usepackage{txfonts}
\usepackage{xcolor}
\usepackage{upgreek}
\usepackage{natbib}
\usepackage{adjustbox}
\usepackage{gensymb}
\usepackage[colorlinks,citecolor=blue,urlcolor=blue,filecolor=blue,linkcolor=blue]{hyperref}
\usepackage{multirow}
\usepackage{amsmath}
\usepackage{tabularx}

\begin{document}

   \title{Tied-array beam flatfielding}

    \author{D. Kuiper\inst{1}
            \and C.\ G.\ Bassa\inst{2}
            \and Z.\ Pleunis\inst{1,2}
            \and J.\ W.\ T.\ Hessels\inst{1,2,3,4}}
    
    \institute{Anton Pannekoek Institute for Astronomy, University of Amsterdam,
               Science Park 904, 1098 XH Amsterdam, The Netherlands\label{api}
               \and
               ASTRON, Netherlands Institute for Radio Astronomy,
               Oude Hoogeveensedijk 4, 7991 PD Dwingeloo, The Netherlands\label{astron}
               \and
               Trottier Space Institute, McGill University,
               3550 rue University, Montréal, QC H3A 2A7, Canada
               \and
               Department of Physics, McGill University,
               3600 rue University, Montréal, QC H3A 2T8, Canada}

   \date{Received \today; accepted \today}

   \abstract
   {Modern multi-element phased-array radio telescopes increasingly employ digital phase coherent beamforming methods to increase their instantaneous field-of-view by filling the primary beam with tens to hundreds of tied-array beams. Since digital beamforming is a linear combination of input antenna voltages to form multiple output tied-array beams, these beams will share similarities in bandpass and gain variations as well as radio frequency interference (RFI). However, most time-domain pipelines for pulsar and fast-transient searches still process each beam separately, rather than using this spatial information to suppress common-mode contamination before the search, so large numbers of RFI-dominated candidates must later be grouped, sifted, and classified.}
   {We exploit the spatial information of multi-beam observations to stabilize bandpasses, suppress red noise and broad-band RFI, and reduce single-pulse false positives, without degrading the detectability of genuine astrophysical signals.}
   {We derive the expected tied-array gain as a function of residual phase dispersion and show how off-target sources converge toward the incoherent limit. Using chi-squared noise statistics, we analyze the impact of dividing a tied-array beam (TAB) by a beam-averaged reference and quantify the smoothing needed for a stable divisor. We then test these predictions using raw LOFAR high-band antenna (HBA) voltages of PSR~B0329+54, simulations, and real LOTAAS survey data of PSR~J0250+5854 processed with and without beam flatfielding.}
   {Once the residual phases decorrelate, outer TABs converge to the incoherent limit set by the effective number of contributing stations, and off-target sources contribute nearly uniform power across beams after primary-beam effects are accounted for. A suitably smoothed null-beam or multi-beam reference provides a very stable divisor, yielding flatter dynamic spectra and equal or higher pulse signal-to-noise ratio than incoherent subtraction. In LOTAAS data for PSR~J0250+5854, beam flatfielding reduces off-dispersion-measure single-pulse triggers by a factor of about $200$ while preserving the folded profile morphology and peak S/N.}
   {Beam flatfielding is a simple post-beamforming operation that requires no heavy computational resources or additional hardware. For current and future multi-beam facilities, especially systems with large phased-array or tied-array mosaics, it offers a direct route to more stable bandpasses, closer-to-Gaussian noise statistics, and order-of-magnitude reductions in single-pulse false positives and downstream classification load at fixed sensitivity.}

   \maketitle

\section{Introduction}
Time-domain radio astronomy has entered a regime where wide fields of view (tens to hundreds of square degrees), high time resolution (ms to tens of $\upmu$s), and large bandwidths (100's to 1000's of MHz) are routinely combined. Facilities such as LOFAR \citep{haa2013}, CHIME \citep{chi2018}, and MeerKAT \citep{mkt2016} already discover large samples of pulsars, fast radio bursts (FRBs), and other fast radio transients, while the SKA \citep{ska2009} will push this further in both sensitivity and survey speed. This capability is increasingly realized through digital phase coherent beamforming methods where signals from individual steerable dishes or stationary antennas/tiles are coherently added to form tens to hundreds of simultaneous tied-array beams per pointing for pulsar and dispersed single-pulse searches \citep[e.g.,][]{ngc2017, cal2017, san2019, raj2022}. These beamformed observing modes deliver unparalleled instantaneous sky coverage, but they also expose a simple fact: imperfections that arise upstream of the beamformer,  including radio-frequency interference (RFI), tend to imprint simultaneously on all beams at a given time and frequency, and must be controlled if we want to exploit the inherent sensitivity of the telescope.

The main limiting factor is no longer thermal noise alone but the combination of structured foregrounds, instrumental bandpass variation, and an increasingly crowded radio spectrum. Modern observatories operate in environments where RFI from communication networks, aircraft, satellites, and digital electronics is both ubiquitous and dynamic \citep[e.g.,][]{off2013, off2015, sat2023, zha2025}. These effects are especially severe at low radio frequencies. For classical dish arrays, part of this environment is physically shadowed by the telescope structure and the horizon mask of the reflector. In contrast, instruments like LOFAR and SKA-Low are built from large numbers of small, fixed antennas in sparse stations, with no mechanical horizon shielding \citep{haa2013, low2021}. Every antenna element ``sees'' most of the visible sky at once, and strong near-horizon transmitters or satellites can illuminate the array through sidelobes and grating lobes in directions that are difficult to model or avoid.

In current pulsar and fast-transient pipelines, RFI is usually handled by masking contaminated parts of the dynamic spectrum, where a dynamic spectrum is the measured intensity as a function of observing time and radio frequency. Standard tools such as \texttt{rfifind} in \textsc{PRESTO} \citep{ransom2002}, zero-DM and median filters \citep{eat2009}, and more recent interquartile range mitigation (IQRM) type algorithms \citep{mor2022} all operate in the time–frequency plane on individual dynamic spectra, excising or masking/flagging time samples or whole channels whose statistics deviate from the local baseline prior to signal candidate detection. Machine-learning methods are now consistently used at the candidate stage to down-select single-pulse and periodicity detections, operating on diagnostic plots or engineered features, but again work on a per-beam basis \citep[e.g.,][]{mor2014,mic2018,aga2020}. These methods are successful at suppressing obvious RFI, but the price is lost bandwidth and exposure time: heavily contaminated channels or even whole sub-bands are commonly flagged out of the analysis. At low frequencies, where broad-band and slowly varying contaminants are common, such excision can significantly reduce the useful bandwidth and inflate the variance of the remaining samples. In periodicity searches, persistent instrumental tones (often called ``birdies'') can likewise produce spurious Fourier peaks that must be masked. Even after these steps, large numbers of spurious single-pulse and periodicity candidates often remain, consuming storage, computing resources, and human or machine classification effort. A small number of approaches try to use spatial redundancy earlier, for example by subtracting an incoherent beam as a common-mode reference to suppress broad-band fluctuations in tied-array data \citep[e.g.,][]{roy2018,price2024}. Nonetheless, beyond such cases and coincidence filters at the candidate stage, the signal processing and RFI excision steps leading up to the search almost exclusively treat each beam independently. Consequently, they fail to exploit the fact that many beams probe nearly the same sky and the same instrumental effects at the same time.

This redundancy follows directly from how phase coherently beamformed tied-array beams (TABs)\footnote{The phase coherent sum of signals from different antennas is sometimes also called the coherent beam or phased-array beam.} are formed \citep[see, e.g.,][]{tho2017}. In a TAB, complex voltages from $N_a$ stations are phase aligned toward a chosen direction on the sky before summation. On boresight, the phase-aligned sum yields an array gain that scales as $N_a^2$. As the source moves away from that direction, residual phase differences accumulate and the coherent sum gradually decorrelates, transitioning toward the incoherent limit where the net gain scales as $N_a$ instead. For an off-boresight point source, the phases at different stations are effectively random, and the expected gain contribution in any given beam approaches the incoherent sum of station powers, modulated only by the smooth primary-beam envelopes and bandpass shapes of the stations. In other words, apart from the beam(s) that happen to be near the source, all other beams see nearly the same average contribution from that source, folded together with the same instrumental bandpass, sky temperature variations, and broad-band RFI.

Seen from this perspective, a multi-beam observation is not simply a collection of independent dynamic spectra. At each frequency–time sample, the power across beams shares a strong, slowly varying multiplicative component: the common bandpass, gain, and sky temperature structure plus any broad-band, near-field RFI that illuminates the whole station. Genuine astronomical variability from a compact source, in contrast, is confined to a small footprint in beam space in the far-field set by the tied-array point-spread function, that is, the angular response pattern of the formed beam to a point source. The same is true in dispersion-measure (DM) space: an FRB or pulsar pulse is detected only over a limited range of trial DMs around its true value, whereas large-scale instrumental drifts appear across many or all DMs.

This observation motivates an analogue of optical flatfielding in beam space: \emph{beam flatfielding}. Instead of treating each TAB in isolation, we use the ensemble of beams to estimate the common multiplicative structure in the data. For each beam and for each time–frequency sample, a mean over the other beams provides a reference that tracks station bandpasses, slow gain drifts, and broad-band RFI that is seen by most beams. Dividing the target beam by this reference removes most of the shared structure while leaving the relative differences between beams intact. For an on-target astrophysical signal, this operation rescales the pulse amplitude but, provided the reference varies only on timescales longer than the pulse width, does not appreciably distort its morphology; for components that are common across beams, it suppresses both their mean level and their variance.

In this paper we formalize this picture and quantify its consequences for pulsar and single-pulse searches with TABs. First, we derive the expected beam response as a function of residual phase dispersion and show explicitly how the coherent array gain transitions to the incoherent limit for off-target sources, including the dependence on the rms phase error. This provides a statistical basis for treating off-target sources and broad-band RFI as approximately uniform contributors across beams. We then define a practical beam flatfielding operator that uses the other beams in a pointing to construct a reference dynamic spectrum for each beam, and we analyze its effect on the per-sample signal-to-noise (S/N) ratio and on the noise statistics after normalization. We validate these predictions and optimize the beam flatfielding strategy -- specifically, the use of dedicated null-beam references -- using raw LOFAR high-band antenna (HBA) voltages of PSR~B0329+54, for which we form tied-array, incoherent, and null beams offline. To test the underlying assumptions on a survey-wide scale, we simulate a LOFAR-style configuration of stations with the actual LOTAAS beam geometry as defined by \citet{san2019} and evaluate how the weighted coherence and normalized power behave across the field of view. These simulations confirm that TABs near the boresight remain fully coherent while outer beams quickly converge to the incoherent regime, and that off-target sources contribute nearly uniform power once the primary-beam envelope is accounted for. Finally, we apply the beam flatfielding procedure to real LOFAR data from LOTAAS, using  the ultra-slow ($P=23.5$\,s) PSR~J0250+5854 \citep{tan2018} as a case study. In this observation, beam flatfielding reduces the number of single-pulse false positives by almost two orders of magnitude, while preserving the folded pulsar profile and improving the statistical stability of the baseline.

The approach developed here is computationally cheap, requires no additional hardware, and is directly applicable to current and future multi-beam surveys with LOFAR, CHIME, and other interferometric radio telescopes. By exploiting the spatial information inherent in multi-beam observations, beam flatfielding complements existing RFI mitigation and calibration schemes, and offers a simple path to more robust transient searches in increasingly crowded radio environments.

In the following, Sect.~\ref{sec:beamforming} reviews Tied-array beamforming and the transition from coherent to incoherent summation. Sect.~\ref{sec:flatfielding} introduces the beam flatfielding operator and derives its statistical properties. Sect.~\ref{sec:stations} and Sect.~\ref{sec:simulations} test the assumptions with station-level data and simulations, and Sect.~\ref{sec:realdata} presents a LOTAAS case study demonstrating the impact on real observations. We conclude with implications for future multi-beam arrays and transient search pipelines. Table~\ref{tab:terminology} summarizes the terminology used throughout this paper; in particular, we use \emph{TAB} for the phased-sum beam and reserve \emph{coherent} and \emph{incoherent} for response regimes or for the incoherent beam itself.

\begin{table*}[t]
    \caption{Terminology used throughout this paper.}
    \label{tab:terminology}
    \centering
    \begin{tabularx}{\textwidth}{l l X}
    \hline\hline
    Term & Abbrev. & Meaning \\
	    \hline
	    Tied-array beam & TAB & Coherently formed beam from phased station voltages. \\
	    Phased-array beamforming & - & Used here synonymously with TABforming: coherent beam formation by phasing and summing voltages. \\
	    Sub-array pointing & SAP & Survey pointing containing a set of TABs. \\
	    Dynamic spectrum & - & Intensity recorded as a function of observing time and radio frequency. \\
    Point-spread function & PSF & Angular response pattern of a formed beam to a point source. \\
    Target beam & - & Beam being flatfielded and evaluated. \\
    Null beam & - & Beam placed near a response minimum to provide a reference with little source power. \\
    Incoherent beam & - & Sum of station powers without phase alignment. \\
    Beam flatfielding & - & Normalization of a beam by a smoothed multi-beam reference. \\
    Null division & - & Beam flatfielding using a null-beam reference. \\
    Incoherent subtraction & - & Subtraction of a scaled incoherent beam used as a comparison method. \\
    \hline
    \end{tabularx}
\end{table*}

\section{Beamforming}
\label{sec:beamforming}

In tied-array or phased-array beamforming, the voltages $V_i$\footnote{Throughout this paper we use the complex-voltage formulation, so $V_i$ denotes a complex-sampled voltage stream. An equivalent beamforming description can be written for real-sampled voltages.} from $N_a$ individual antennas or stations are weighted with complex phasors $\omega_i$ and summed
\begin{equation}
    b = \sum_{i=1}^{N_a} \omega_i V_i,
    \label{eq:bf_single}
\end{equation}
to form the complex voltage of a TAB $b$. For a plane wave incident from a direction with direction cosines $(l,m,n)$, with $n=\sqrt{1-l^2-m^2}$, the geometric phase at station $i$ with coordinates $(u_i,v_i,w_i)$ in units of wavelength is
\begin{equation}
    \phi_i = 2\pi (u_i l + v_i m + w_i (n-1)),
\end{equation}
and the corresponding beamforming weight is $\omega_i = \exp(-\mathrm{i}\phi_i)$. When multiple beams $b_k$ are formed, Eq.~\ref{eq:bf_single} generalises to
\begin{equation}
    b_k = \sum_{i=1}^{N_a} \omega_{ik} V_i
        = \sum_{i=1}^{N_a} \exp\left[-2\pi \mathrm{i} (u_i l_k + v_i m_k + w_i (n_k-1))\right] V_i,
    \label{eq:bf_multiple}
\end{equation}
where each beam $k$ has its own set of direction cosines $(l_k,m_k,n_k)$ and thus its own weights $\omega_{ik}$. Throughout this paper, $(u_i,v_i,w_i)$ and $(l_k,m_k,n_k)$ are real-valued coordinates, whereas $V_i$, $b_k$, and $\omega_{ik}$ are complex. When useful we denote the full set of beamforming weights for beam $k$ by the complex vector $\boldsymbol{\omega}_k$, but otherwise work component-wise.

For a narrowband source with complex voltage at station $i$ written as
\begin{equation}
    V_i = A\,\exp(\mathrm{i}\psi_i),
\end{equation}
with real amplitude $A$ and total phase $\psi_i$, the beam voltage becomes
\begin{equation}
    b_k = A \sum_{i=1}^{N_a} \exp\left[\mathrm{i}(\psi_i - \phi_{ik})\right]
        = A \sum_{i=1}^{N_a} \exp(\mathrm{i}\delta_{ik}),
\end{equation}
where
\begin{equation}
    \delta_{ik} \equiv \psi_i - \phi_{ik}
\end{equation}
is the residual phase between the true source direction and the beam pointing. The detected beam power is
\begin{equation}
\label{eq: beam_power}
    I_k = |b_k|^2
        = |A|^2 \left|\sum_{i=1}^{N_a} \exp(\mathrm{i}\delta_{ik})\right|^2.
\end{equation}

\begin{figure}[t]
    \centering
    \includegraphics[width=\linewidth]{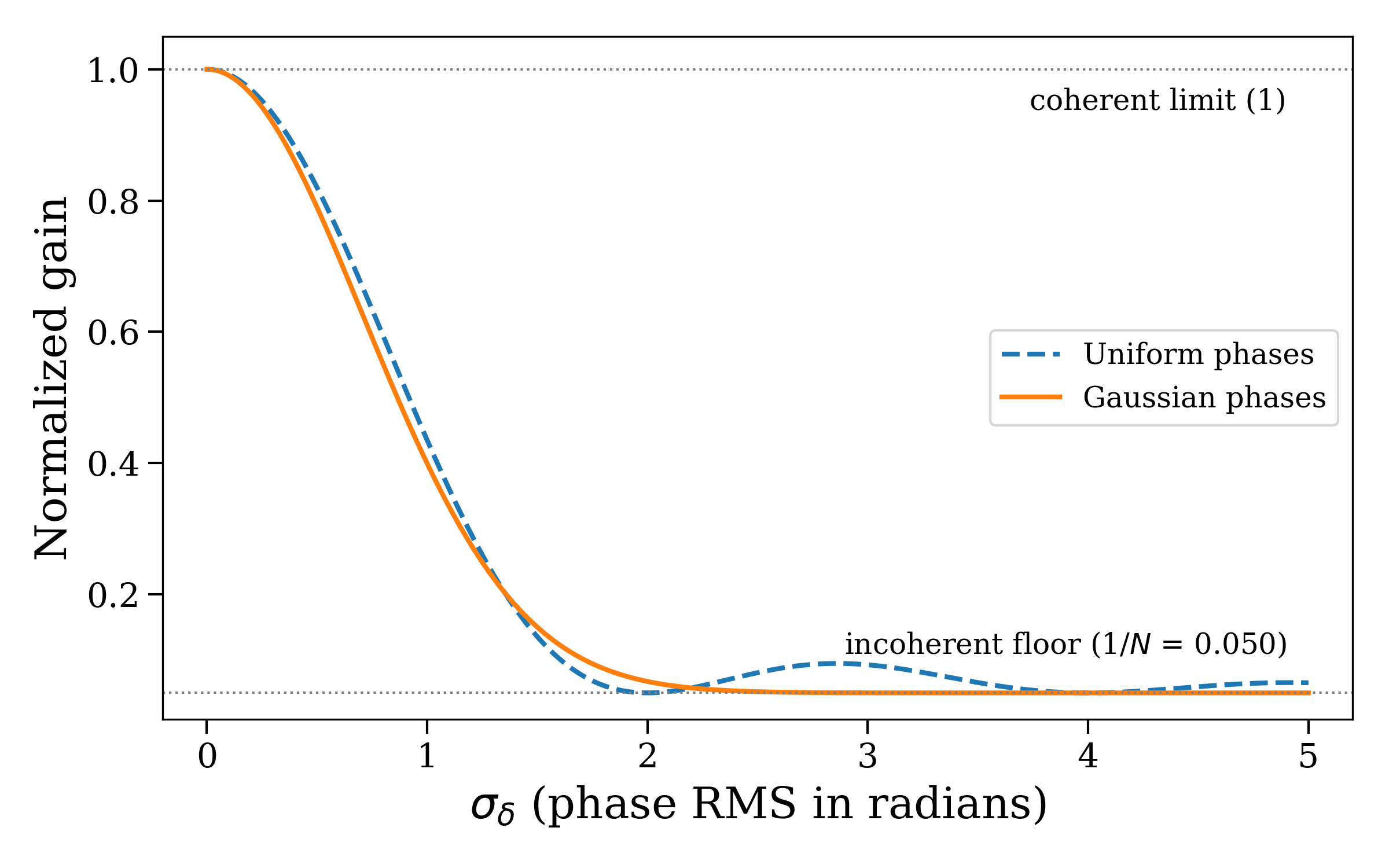}
    \caption{Normalized array gain as a function of rms phase error $\sigma_\delta$ for $N_a=20$, comparing uniform (dashed blue) and Gaussian (solid orange) phase distributions.
    The response is normalized by $N_a^2$ to show the transition from the fully coherent limit ($=1$) to the incoherent limit ($=1/N_a$).}
    \label{fig:coh_incoh_transition}
\end{figure}

To understand how the array response changes as a source moves away from the the beam centre, we are interested in the expected value $\mathbb{E}[I_k]$ as a function of the residual phase dispersion. Splitting Eq.~\ref{eq: beam_power} into diagonal and off-diagonal terms and averaging over random residual phases gives
\begin{equation}
    \mathbb{E}[I_k]
    = |A|^2 \Big[ N_a + N_a(N_a-1)\,\rho(\sigma_\delta) \Big],
    \label{eq:Ik_expect_general}
\end{equation}
where
\begin{equation}
    \rho(\sigma_\delta)
    \equiv \left|\mathbb{E}\left[\exp(\mathrm{i}\delta)\right]\right|^2
\end{equation}
is the complex phase coherence and $\sigma_\delta$ is the rms of the residual phase distribution. For a simple example with $\delta$ uniformly distributed over the interval $[-a,a]$ one finds
\begin{equation}
    \mathbb{E}\left[\exp(\mathrm{i}\delta)\right] = \frac{\sin a}{a},
\end{equation}
and with $\sigma_\delta^2=a^2/3$ this gives
\begin{equation}
    \rho(\sigma_\delta)
    = \left(\frac{\sin(\sqrt{3}\,\sigma_\delta)}{\sqrt{3}\,\sigma_\delta}\right)^2.
\end{equation}
If the residual phases instead follow a Gaussian distribution, the coherence factor simplifies to $\rho(\sigma_\delta) = \exp(-\sigma_\delta^2)$.\\
Normalising Eq.~\ref{eq:Ik_expect_general} by the fully coherent gain $N_a^2 |A|^2$ yields the dimensionless array gain
\begin{equation}
    G(\sigma_\delta)
    \equiv \frac{\mathbb{E}[I_k]}{N_a^2 |A|^2}
    = \frac{1}{N_a} + \left(1-\frac{1}{N_a}\right)\rho(\sigma_\delta),
\end{equation}
which decreases from unity at $\sigma_\delta=0$ to $1/N_a$ in the incoherent limit where the residual phases decorrelate. Figure~\ref{fig:coh_incoh_transition} illustrates this transition for $N_a=20$.

The key point for this work is that once the residual phase dispersion is large enough that $\rho(\sigma_\delta)\ll 1$, the expected power from a point source in any given beam approaches the incoherent limit
\begin{equation}
    \mathbb{E}\{I_k\} \rightarrow N_a\,|A|^2,
\end{equation}
independent of the precise beam pointing. Apart from the smooth primary beam envelope encoded in the station amplitudes $A_i$, an off-target source then contributes approximately the same average detected power to every simultaneously formed beam. The same is true for broad-band near-field RFI that illuminates all stations. Consequently, the per-beam dynamic spectra share a common large-scale multiplicative structure set by the station bandpasses, sky temperature, and any broad-band RFI. This motivates the beam flatfielding developed in Sect.~\ref{sec:flatfielding}, where an average over the other beams at each $(\nu,t)$ is used as a reference estimate of this common structure. The detailed derivation of Eq.~\ref{eq:Ik_expect_general} is given in Appendix~\ref{app:array_gain}.

\section{Beam flatfielding}
\label{sec:flatfielding}

Section~\ref{sec:beamforming} showed that as the residual phase dispersion
$\sigma_\delta$ increases, the tied-array response transitions from the fully
coherent to the incoherent limit. In this regime, off-target sources and broad-band
RFI contribute nearly the same average detected power to every simultaneously
formed beam, apart from a smooth primary-beam envelope and slowly varying
instrumental structure. This implies that at a given frequency and time, the set
of TABs share a common large-scale multiplicative pattern on top of
which any genuine astronomical signal appears only in a small subset of beams.
The goal of beam flatfielding is to remove this shared structure while preserving
on-target signals.

\subsection{Noise statistics in brief}

In pure noise, the complex voltage at a single station and single
polarization can be written as
\begin{equation}
    V = X + \mathrm{i}Y,
\end{equation}
with $X$ and $Y$ independent Gaussian variables with zero mean and equal
variance. The instantaneous power in one polarization is proportional to
$|V|^2 = X^2 + Y^2$, and Stokes~$I$ for both polarizations is the sum of four
squared Gaussian components. For LOFAR these two polarimetric channels are the
orthogonal linear receptor voltages, so Stokes~$I$ is formed by summing the
detected powers of the two linear polarizations. A single time–frequency sample
of station Stokes~$I$ therefore follows a scaled $\chi^2_4$ variate, where the subscript denotes the
four degrees of freedom.

For a TAB, complex voltages from $N_a$ stations are summed
before detection. The sum of Gaussian voltages is again Gaussian, so the coherent
beam voltage in each polarization is still complex Gaussian and the coherent
Stokes~$I$ sample is again a scaled $\chi^2_4$ variate, just with a larger
overall variance due to the coherent gain.

For an incoherent beam, powers are formed per station and then summed. Each
station power is a scaled $\chi^2_4$ variate, and the incoherent sum of $N_a$
stations is therefore a scaled $\chi^2_{4N_a}$ variate, with $4N_a$ degrees of
freedom. At the level of a single
time–frequency sample, the tied-array and incoherent beams thus have different
noise distributions: a low-degree-of-freedom $\chi^2_4$ for the tied-array case,
and a high-degree-of-freedom $\chi^2_{4N_a}$ for the incoherent sum.

In practice, intensities are averaged in time and frequency before further
processing. If a single sample follows a $\chi^2_k$ distribution (up to an
overall scale), the sum of $M$ independent samples follows $\chi^2_{kM}$ and the
average has mean $\mu \propto k$ and variance $\sigma^2 \propto 2k/M$. The
fractional rms scales as
\begin{equation}
    \frac{\sigma}{\mu} \propto \sqrt{\frac{2}{kM}}.
\end{equation}
For the TAB, $k=4$ and
\begin{equation}
    \left(\frac{\sigma}{\mu}\right)_{\mathrm{tab}}
    \approx \sqrt{\frac{1}{2M}},
\end{equation}
while for the incoherent sum, $k=4N_a$ and
\begin{equation}
    \left(\frac{\sigma}{\mu}\right)_{\mathrm{incoh}}
    \approx \sqrt{\frac{1}{2N_a M}}.
\end{equation}
At fixed averaging scale $M$, the incoherent beam therefore has smaller
fractional noise fluctuations by a factor of order $\sqrt{N_a}$ and a
distribution that is already very close to Gaussian. A beam-averaged reference that
is built from many such beams will be extremely stable. The full derivation of
these results is given in Appendix~\ref{app:noise_stats}.

\subsection{Beam flatfielding operator and effective S/N}

Let $I_k(\nu,t)$ be the detected Stokes~$I$ power in beam $k$ at frequency $\nu$
and time $t$. We model it as
\begin{equation}
\label{eq:I_model_compact}
    I_k(\nu,t) = G_k(\nu,t)\,[\,S_k(\nu,t) + N_k(\nu,t)\,] + C(\nu,t),
\end{equation}
where $G_k$ captures slowly varying multiplicative structure (bandpass, gain,
sky temperature), $C$ is a common additive term across beams, $S_k$ is the
astrophysical signal, and $N_k$ is zero-mean noise with statistics given above.
All quantities in Eq.~\ref{eq:I_model_compact} are real-valued detected powers
or intensity-like terms.

For a given beam $k$ we define a beam-averaged reference by averaging over the
other $M_{\mathrm{beam}}-1$ beams,
\begin{equation}
\label{eq:ref_mean_compact}
    \widehat{B}_k(\nu,t)
    = \frac{1}{M_{\mathrm{beam}}-1}
      \sum_{\substack{j=1 \\ j\neq k}}^{M_{\mathrm{beam}}} I_j(\nu,t),
\end{equation}
and obtain the flatfielded dynamic spectrum
\begin{equation}
\label{eq:ff_main_compact}
    \widetilde{I}_k(\nu,t) = \frac{I_k(\nu,t)}{\widehat{B}_k(\nu,t)}.
\end{equation}
\footnote{Here and below, hats denote reference estimates constructed from other beams in the detected-power domain. Overbars denote frequency-averaged quantities, and tildes denote transformed quantities such as flatfielded spectra or robust baseline estimates.}

Since $G_k(\nu, t)$ is attenuated by the position of this TAB within the response of the primary beam, we divide by the beam-averaged reference dynamic spectrum as opposed to subtracting it from the target beam. This is in contrast to the method by \citet{roy2018} where the incoherent sum of input antennas is subtracted from the target beam, since in that case the incoherent beam includes the full response of the primary beam.

For an on-source sample in beam $k_0$ we write
\begin{equation}
    I_{k_0} = \mu_{\mathrm{tab}} + S + n,\qquad
    \widehat{B}_{k_0} = \mu_B + \varepsilon,
\end{equation}
where $\mu_{\mathrm{tab}}$ is the mean noise level in the TAB, $S$ is the
signal contribution, $n$ is zero-mean tied-array-beam noise with variance
$\sigma_{\mathrm{tab}}^2$, and $\varepsilon$ is the zero-mean fluctuation of
the reference with variance $\sigma_B^2$. The flatfielded sample is
\begin{equation}
    \widetilde{I}_{k_0} = \frac{\mu_{\mathrm{tab}} + S + n}{\mu_B + \varepsilon}.
\end{equation}
Expanding to first order in $\varepsilon/\mu_B$ and using that $n$ and
$\varepsilon$ are independent (beam $k_0$ is excluded from the reference) gives
\begin{equation}
    \mathrm{Var}(\widetilde{I}_{k_0})
    \approx \frac{\sigma_{\mathrm{tab}}^2}{\mu_B^2}
    + \frac{(\mu_{\mathrm{tab}}+S)^2}{\mu_B^4}\,\sigma_B^2.
\end{equation}

Using the chi-squared scaling above, the fractional variance of the tied-array-beam
noise is of order $1/(2M)$ while the fractional variance of the reference is of
order $1/[2N_a M (M_{\mathrm{beam}}-1)]$. Their ratio is approximately
\begin{equation}
    \frac{(\mu_{\mathrm{tab}}+S)^2 \sigma_B^2}{\mu_B^2 \sigma_{\mathrm{tab}}^2}
    \sim \frac{1}{N_a (M_{\mathrm{beam}}-1)},
\end{equation}
which is much smaller than unity for realistic arrays. The variance budget of
the flatfielded beam is therefore dominated by the tied-array-beam noise, not by
the uncertainty in the reference, and the instantaneous S/N loss from dividing
by a noisy reference is negligible. The main effect of beam flatfielding is instead
to remove the large-scale multiplicative structure $G_k$ and $C$, which shrinks
non-Gaussian wings in the noise distribution and reduces the number of false
positives at a fixed S/N threshold.

The detailed algebra behind these scalings is given in
Appendix~\ref{app:noise_stats}. In Sect.~\ref{sec:stations} we test these
expectations directly using raw voltage data of PSR~B0329+54.

\section{Optimizing beam flatfielding using voltage data}
\label{sec:stations}

To validate the statistical framework in Sect.~\ref{sec:flatfielding} and to
optimize the choice of reference beam(s), a short ($\sim$0.5\,s) raw-voltage observation  of the
bright pulsar PSR~B0329+54 with the LOFAR HBAs was used. In this dataset,
complex voltages from 12 24-tile sub-stations on the Superterp were recorded over
20 contiguous subbands (each 195.3125\,kHz wide), spanning 139.16-143.05\,MHz, at the native 5.12\,$\upmu${s} sampling,
without real-time TABforming (though the individual sub-station beams are already formed). This allows tied-array, incoherent,
and dedicated null beams to be formed offline from exactly the same input
data.

\subsection{Observation and beamforming setup}
\label{subsec:stations_setup}

The raw voltages from each sub-station beam are first channelised into fine channels
the COBALT correlator \cite{bro2018}, yielding per-station spectra
$V_{x,i}(t,\nu)$ and $V_{y,i}(t,\nu)$ in two orthogonal polarisations. For
a given sky direction with direction cosines $(l_k,m_k,n_k)$, tied-array
beams are formed using the phased-array sum
(Eq.~\ref{eq:bf_multiple}),
\begin{equation}
    B_{x,k}(t,\nu) = \sum_i \omega_i^*(\nu;l_k,m_k,n_k)\,V_{x,i}(t,\nu),
\end{equation}
and similarly for $B_{y,k}$. Stokes~$I$ for beam $k$ is then
\begin{equation}
    I_k(t,\nu) = |B_{x,k}|^2 + |B_{y,k}|^2.
\end{equation}
In parallel an incoherent beam is formed directly from the station powers,
\begin{equation}
    I_{\mathrm{incoh}}(t,\nu) = \sum_i \bigl(|V_{x,i}|^2 + |V_{y,i}|^2\bigr).
\end{equation}

All beams are then brought to the time-frequency resolution used in the
comparison below and dedispersed in the time domain to the known dispersion measure of
PSR~B0329+54. From the resulting dynamic spectra an on-pulse window is defined
around the brightest pulse in the dedispersed time series and an off-pulse
window is taken from a region free of obvious signal. These windows are used
consistently in all S/N and on–off power measurements below.

\subsection{Target, null, and incoherent beams}
\label{subsec:stations_beam_types}

Because the beamforming weights are computed offline, beams can be placed at
arbitrary positions relative to the sub-station beam pattern. At the band
centre of 141.1\,MHz, this leaves substantial freedom: for this pointing the
12-sub-station Superterp TAB has an FWHM of about 0\fdg46, whereas each HBA
sub-station beam has an FWHM of about 5\fdg3 in the same direction, so many
TABs can be placed within a single sub-station-beam FoV \citep{haa2013}. In this test a
one-dimensional cut is constructed by defining a grid of trial directions
$\mathrm{d}\alpha_k$ along right ascension at fixed declination, centred on the
known position of PSR~B0329+54. For each grid point, a TABforming
weight vector $\boldsymbol{\omega}_k$ is computed, so that a synthetic TAB can be formed
at every offset.

From this set the following beams are identified:

\begin{itemize}
  \item \emph{Target beam:} the TAB
  steered to the kown pulsar position. This is the
  beam on which beam flatfielding performance is evaluated.

  \item \emph{Incoherent reference:} the incoherent sum $I_{\mathrm{incoh}}$,
  which has stable, high-degree-of-freedom noise (Appendix~\ref{app:noise_stats})
  but contains the full on-axis pulsar power.

  \item \emph{Null beams:} a set of TABs located near
  predicted minima of the sub-station beam response. A normalized tied-array
  response profile along the cut is computed from the weights alone using
  the quadratic form
  \begin{equation}
      R(\mathrm{d}\alpha_k) \propto
      \left| \sum_{i,\nu} \omega_i^*(\nu;0)\,\omega_i(\nu;\mathrm{d}\alpha_k) \right|^2,
  \end{equation}
  which measures the overlap between the central beam and the beam at offset
  $\mathrm{d}\alpha_k$. The beam with the highest response defines the target
  beam; the 48 beams with the smallest normalized response, excluding this
  central peak, are selected as null beams.
\end{itemize}

\begin{figure}[t]
    \centering
    \includegraphics[width=\linewidth]{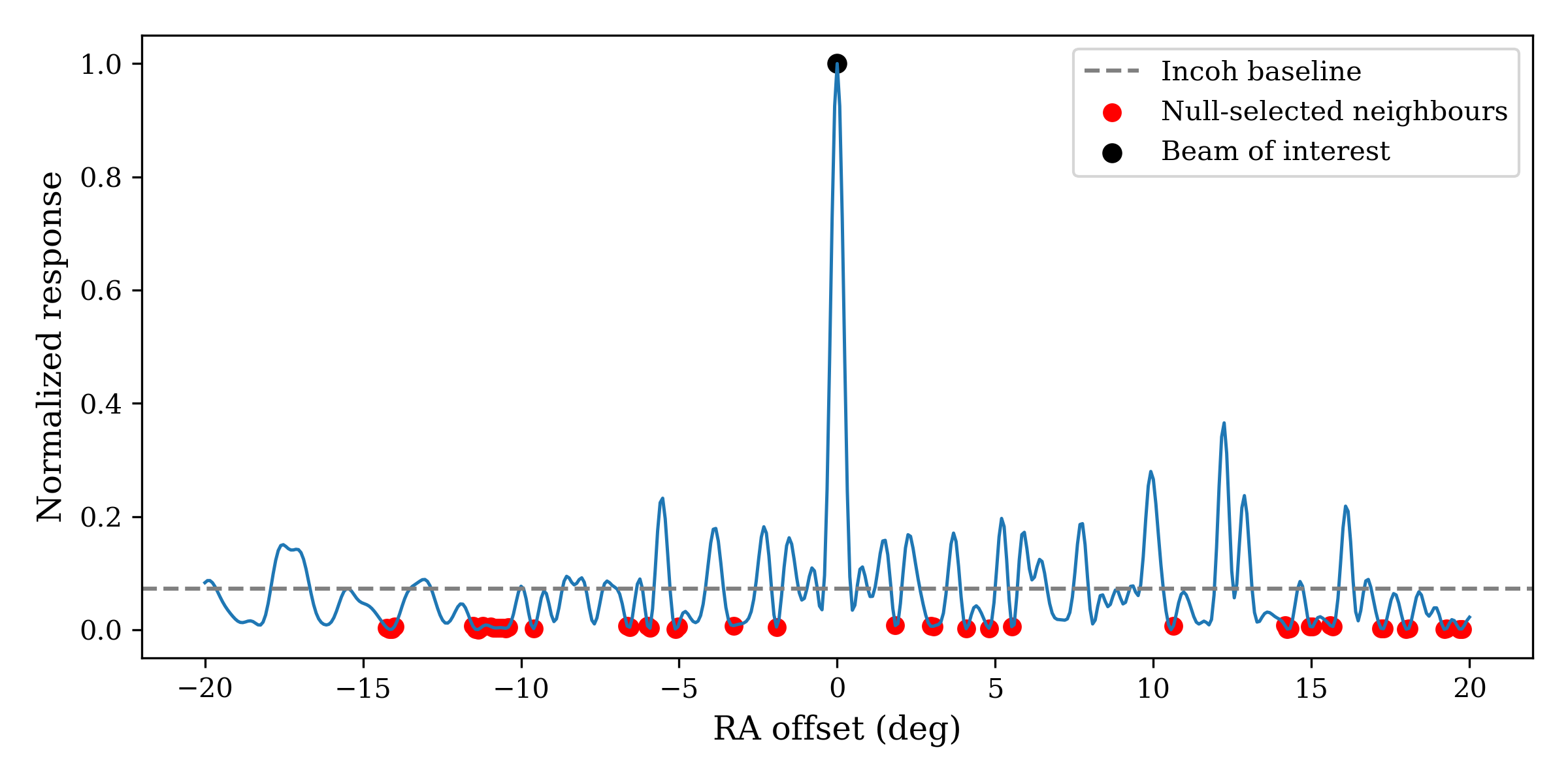}
    \caption{Normalized to unity tied-array response along a one-dimensional cut in right
    ascension through the PSR~B0329+54 pointing. The central maximum marks the
    target beam. Red dots indicate the null beams with the lowest
    normalized response, which are used to construct the null-beam reference.
    The dotted horizontal line marks the incoherent-power reference level
    $1/N_a = 1/12 \approx 0.083$ for this 12-sub-station Superterp setup.}
    \label{fig:null_response}
\end{figure}

Figure~\ref{fig:null_response} shows the resulting response profile along the
cut. The central peak corresponds to the target beam; the coloured markers
indicate the 48 null beams used to build the null-beam reference. These
null beams sit in the sidelobe minima of the station response and
therefore have strong suppression of the on-axis pulsar, while remaining
sufficiently close in angle to share the same bandpass, gain drifts, and
broad-band RFI as the target beam.

For arrays with flexible tied-array backends this suggests a straightforward
survey-design choice: in addition to the main science tiling, one or more
beams per pointing can be deliberately placed near predicted minima of the
sub-station beam pattern to act as dedicated null-beam streams for beam flatfielding.
In real-time processing these null locations would typically be precomputed
from the beam model or beamforming weights for each pointing, rather than inferred from the data, and
the null list can be updated as the pointing changes.

\subsection{Dynamic spectra and beam flatfielding performance}
\label{subsec:stations_dynamics}

Dynamic spectra are then constructed for the target beam, the average of the 48
null beams, and the incoherent beam. To match typical search-pipeline settings
these spectra are smoothed by block averaging in time. The null-beam average
and the incoherent beam are smoothed more strongly than the target beam, so
that their thermal noise is well below the intrinsic noise of the target beam
while still tracking the same large-scale multiplicative structure.

The following processed spectra are formed:

\begin{enumerate}
  \item a null-division spectrum,
  \begin{equation}
      I_{\mathrm{null,div}}(t,\nu)
      = \frac{I_{\mathrm{tar}}(t,\nu)}{\widehat{I}_{\mathrm{null}}(t,\nu)},
  \end{equation}
  where $\widehat{I}_{\mathrm{null}}$ is the smoothed average of the
  48 null beams;

  \item an incoherent-subtraction spectrum,
  \begin{equation}
      I_{\mathrm{incoh,sub}}(t,\nu)
      = I_{\mathrm{tar}}(t,\nu) - \alpha\,I_{\mathrm{incoh}}(t,\nu),
  \end{equation}
  where $\alpha$ is chosen from off-pulse samples to match the mean level of
  the incoherent reference to that of the target beam.
\end{enumerate}

\begin{figure*}[t]
    \centering
    \includegraphics[width=\textwidth]{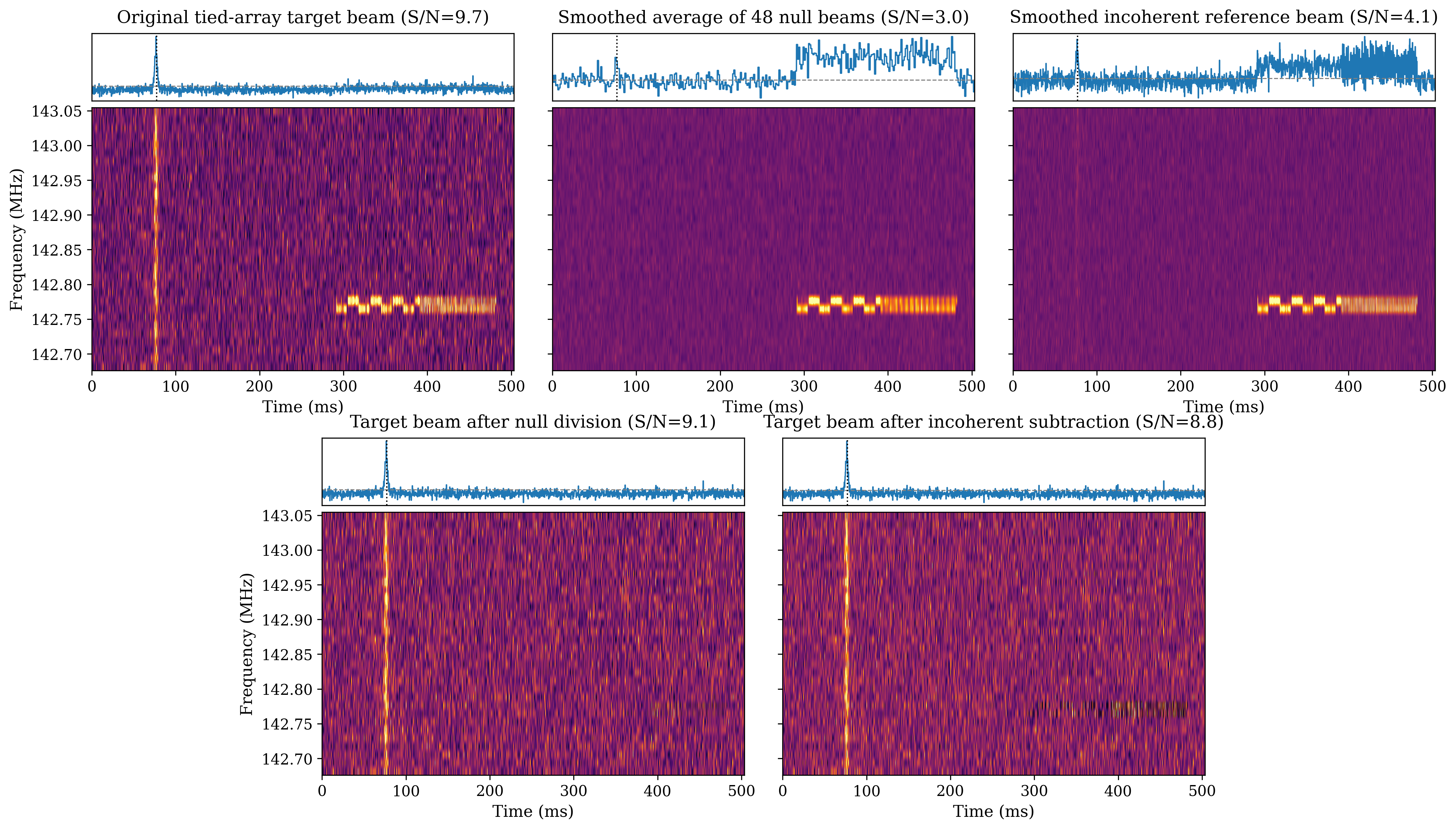}
    \caption{Single-pulse dynamic spectra and time series for PSR~B0329+54 after
    dedispersion to DM$=26.75$~pc~cm$^{-3}$. The x-axis shows time across the
    full displayed segment ($503$~ms), which contains one bright pulse. Top
    row: smoothed target beam (left), smoothed average of the 48 null beams
    (centre), and smoothed incoherent beam (right). Bottom row: target beam
    after null division (left) and incoherent subtraction (right).
    In each panel the lower sub-plot shows the dynamic spectrum and the upper
    sub-plot the frequency-averaged time series with the robust S/N estimate of
    Eq.~\ref{eq:robust_snr}. The
    plotted frequency range is $142.676$-$143.054$~MHz. The target-beam panels
    use a time/frequency resolution of $0.328$~ms and $12.2$~kHz; the null-beam
    average and incoherent reference are smoothed to $1.31$~ms and $12.2$~kHz
    before applying beam flatfielding. For the configuration shown here, null-division gives
    a slightly higher robust S/N than incoherent subtraction ($9.1$ versus
    $8.8$) and suppresses the broad RFI patch in the lower-right more cleanly.
    As shown below, the relative performance depends on the amount of
    null-beam averaging and smoothing.}
    \label{fig:dyn_spectra}
\end{figure*}

Figure~\ref{fig:dyn_spectra} shows the resulting five-panel diagnostic plot.
The top row presents the smoothed target beam, the smoothed average of the 48
null beams, and the smoothed incoherent beam. The bottom row shows the
corresponding null-division spectrum (left) and incoherent-subtraction spectrum
(right). For each panel the lower sub-plot is the dynamic
spectrum and the upper sub-plot shows the frequency-averaged time series and a
robust S/N estimate based on the median and median absolute deviation in the
off-pulse window. More explicitly, for a dynamic spectrum $I(t,\nu)$ we define the
frequency-averaged time series
\begin{equation}
    \bar{I}(t) = \frac{1}{N_\nu}\sum_\nu I(t,\nu),
\end{equation}
the off-pulse baseline
\begin{equation}
    \tilde{I}_{\mathrm{off}} =
    \mathrm{med}_{t \in W_{\mathrm{off}}}\bar{I}(t),
\end{equation}
and the corresponding robust rms estimate
\begin{equation}
    \sigma_{\mathrm{off,MAD}} =
    1.4826\,\mathrm{med}_{t \in W_{\mathrm{off}}}
    \left|\bar{I}(t) - \tilde{I}_{\mathrm{off}}\right|.
\end{equation}
We then define
\begin{equation}
    \mathrm{S/N}_{\mathrm{rob}}
    = \frac{\max_{t \in W_{\mathrm{on}}}\bar{I}(t) -
    \tilde{I}_{\mathrm{off}}}
    {\sigma_{\mathrm{off,MAD}}},
    \label{eq:robust_snr}
\end{equation}
where $\sigma_{\mathrm{off,MAD}}$ is the Gaussian-equivalent rms estimated from
the median absolute deviation (MAD) of the off-pulse samples. This is the S/N
quoted in the panel titles of Fig.~\ref{fig:dyn_spectra}.

Both reference-based methods suppress the broad RFI patch in the lower-right corner
of the target dynamic spectrum. With the modest smoothing used in
Fig.~\ref{fig:dyn_spectra}, the null-division product yields a slightly higher
robust S/N than incoherent subtraction ($9.1$ versus $8.8$ in the panel
titles) while also removing the broad RFI patch more completely. In the
incoherent-subtraction case, the scaled incoherent reference contributes about
$13\%$ of the target-beam pulse peak at the pulse maximum and about $10\%$ of
the integrated on-window pulse fluence, so that this fraction of the source
power is self-subtracted together with the common-mode background. The
null-beam average, by construction, carries almost no pulsar power and
therefore avoids this bias, but its advantage depends on the reference being
made sufficiently stable by averaging over enough beams and/or time samples. As
the null-beam averaging is varied, the reference noise drops and the balance
between the two methods shifts.

For fluence measurements this distinction matters. In the division case,
Eq.~\ref{eq:ff_main_compact} shows that, if the reference varies only slowly
across the pulse window, the flatfielded pulse is approximately the original
pulse multiplied by $1/\widehat{B}_k$: its width and morphology are preserved,
but its amplitude and integrated fluence in the flatfielded units are rescaled
by the local reference level and are therefore not directly physical. In the
subtraction case, the use of an incoherent reference can remove a small fraction
of the source power itself, biasing the measured fluence low. In practice we
therefore regard beam flatfielded products as search products: they are useful
for detection, windowing, and RFI suppression. Even if a particular
flatfielding choice were to incur a modest reduction in peak S/N relative to an
alternative, that trade-off can still improve practical detectability because
orders-of-magnitude fewer false positives reduce the chance that a genuine
astrophysical event is buried in the candidate background or lost during
downstream sifting and classification. Any reported fluence should therefore be
measured from the original calibrated target-beam dynamic spectrum, using the
flatfielded detection only to define the event window and masks.

This qualitative behaviour is exactly what is expected from the
$\chi^2$-based noise arguments in Sect.~\ref{sec:flatfielding} and
Appendix~\ref{app:noise_stats}. To make this connection explicit, we now look
directly at the intensity distributions and the dependence on smoothing
parameters.

\subsection{Noise stability and smoothing requirements}
\label{subsec:stations_noise}

A primary concern with division-based beam flatfielding is the stability of the
denominator. As discussed in Appendix~\ref{app:noise_stats}, a single coherent
null beam at native resolution follows a scaled $\chi^2_4$ distribution (or an
exponential for intensity), which has a high probability density near zero.
Using such a noisy reference directly in the denominator generates extreme
outliers in the flatfielded data.

\begin{figure}[t]
    \centering
        \includegraphics[width=\linewidth]{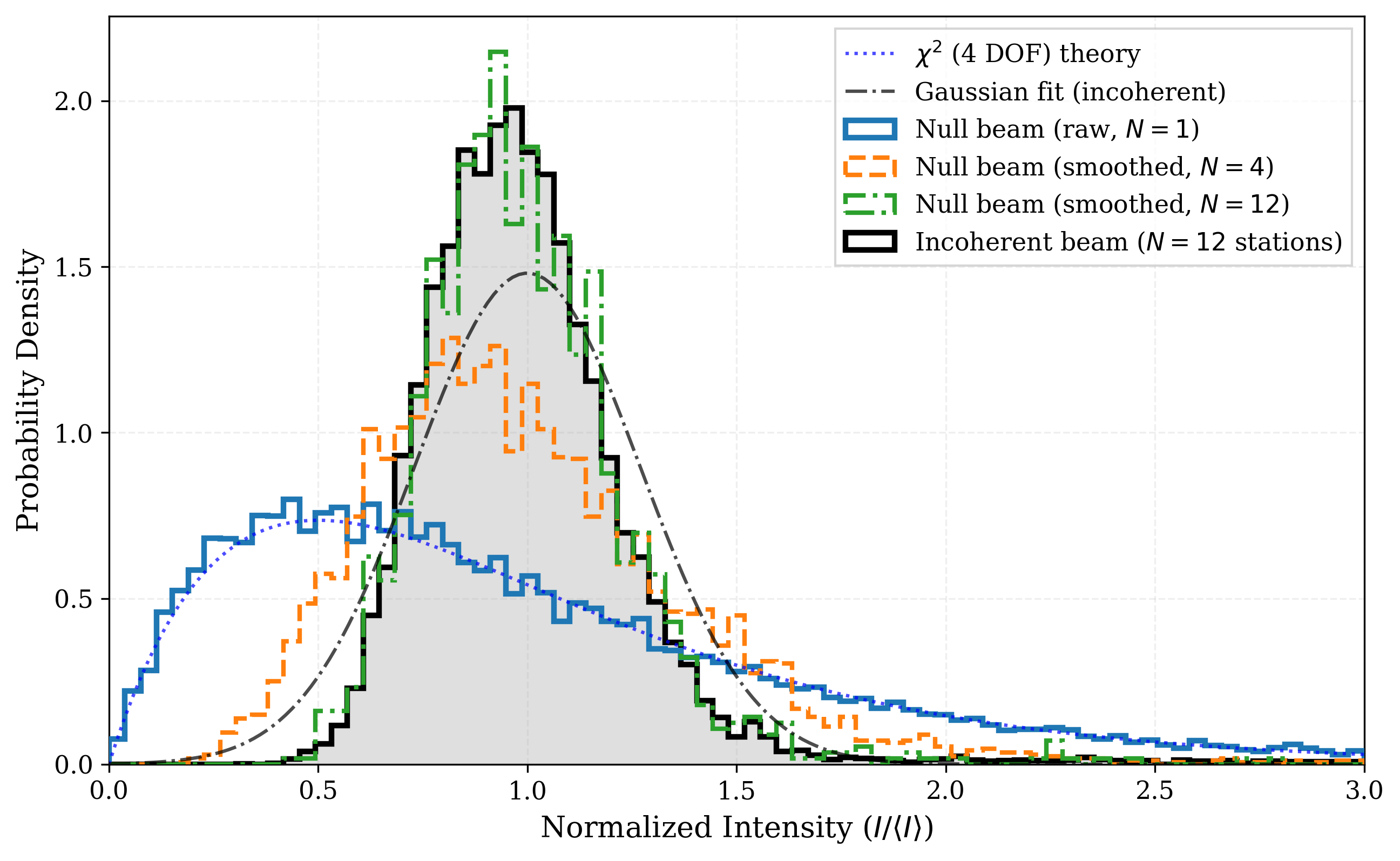}    
        \caption{Probability density functions of normalized intensity for null and incoherent beams measured in off-pulse regions. The raw null beam (blue solid) follows the expected $\chi^2$ distribution with 4 degrees of freedom, showing a wide tail and enhanced probability near zero that makes it unsuitable as a divisor. Temporal smoothing by $N_t=4$ (orange dashed) and $N_t=12$ (green dash-dot) samples progressively narrows the distribution via the Central Limit Theorem. The incoherent beam (gray filled) sums 12 independent sub-stations and exhibits a naturally narrow Gaussian-like distribution; dotted lines show the corresponding theoretical curves. Smoothing by $N_t=12$ reduces the null-beam variance by $2.8\times$, bringing it close to the incoherent case.}
    \label{fig:stats_hist}
\end{figure}

Figure~\ref{fig:stats_hist} compares the normalized intensity distribution of a single null beam to that of the same beam after increasing amounts of temporal smoothing, and to the incoherent reference. The raw null beam (blue solid) closely follows the theoretical $\chi^2$ distribution with 4 degrees of freedom, showing the expected exponential tail and enhanced probability at low
intensities that makes it unsuitable for safe division. After smoothing by $N_t=4$ (orange dashed), the distribution narrows significantly. With $N_t=12$ (green dash-dot), it converges toward the incoherent beam (gray filled), which sums 12 independent stations and naturally exhibits a narrow Gaussian-like distribution; full convergence requires either more smoothing or averaging multiple null beams.

To quantify when null division actually outperforms incoherent subtraction, the
S/N of the pulsar was measured for a grid of smoothing factors and numbers of
null beams. For each point in this grid the ratio
\begin{equation}
    \mathcal{R}_{\mathrm{S/N}} =
    \frac{\mathrm{S/N}_{\mathrm{null}}}{\mathrm{S/N}_{\mathrm{incoh}}}
\end{equation}
was computed, where $\mathrm{S/N}_{\mathrm{null}}$ is the peak S/N in the
null-division spectrum and $\mathrm{S/N}_{\mathrm{incoh}}$ is the peak S/N
after incoherent subtraction with the same total averaging.

\begin{figure*}[t]
    \centering
        \includegraphics[width=0.8\textwidth]{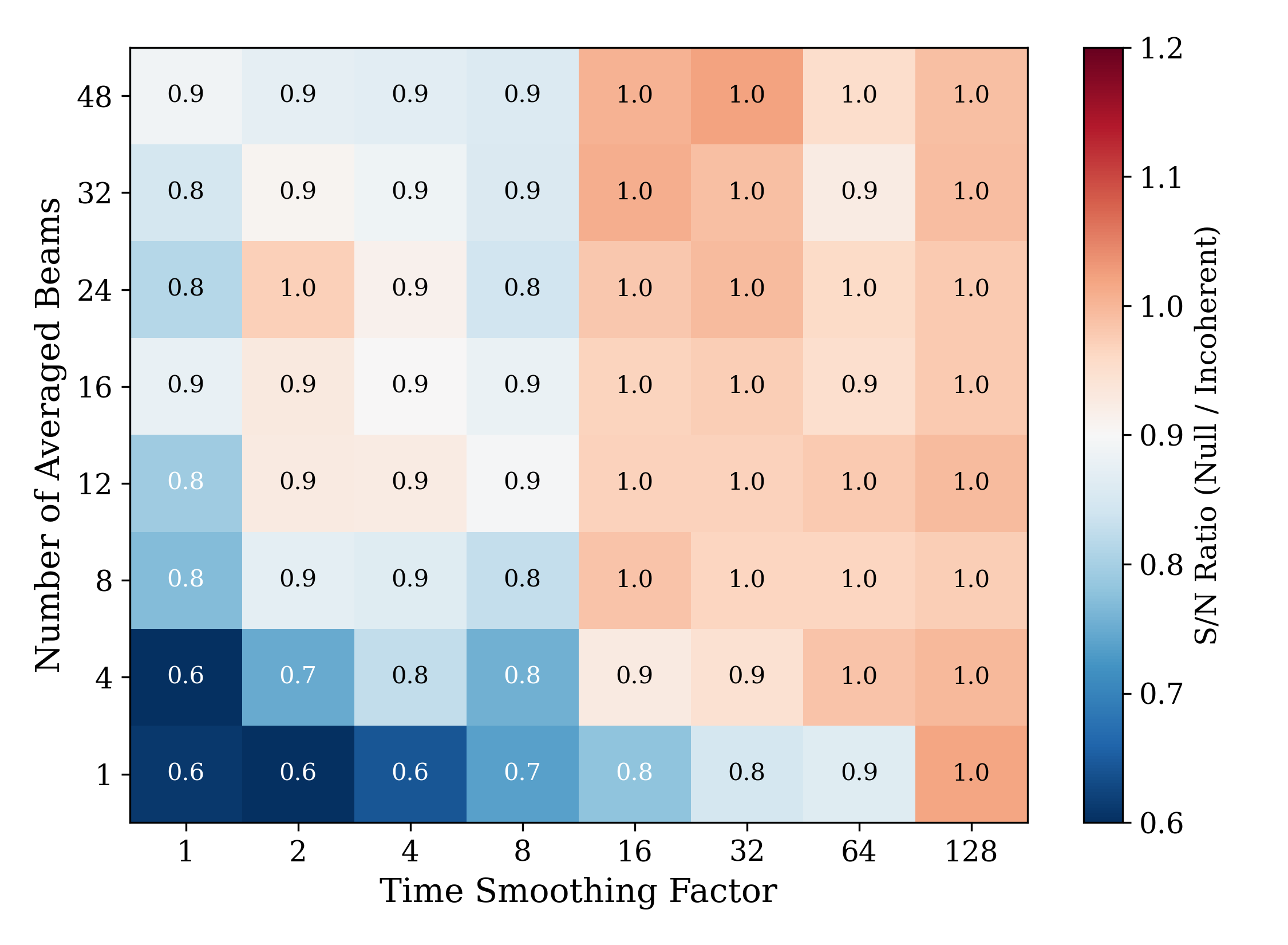}
    \caption{Null-division vs.\ incoherent-subtraction regimes. The colour scale shows
    the ratio of pulsar S/N obtained with null division to that
    obtained with incoherent subtraction, as a function of null-beam smoothing
    factor and the number of null beams averaged. Blue indicates higher
    S/N for incoherent subtraction at low averaging; as smoothing and the null
    beam count increase, the ratio rises toward and above unity because
    null-division avoids self-subtraction and better suppresses broad RFI.}
    \label{fig:regime_map}
\end{figure*}

The result is summarised in Fig.~\ref{fig:regime_map}. The colour scale shows
$\mathcal{R}_{\mathrm{S/N}}$ as a function of the null-beam smoothing factor
and the number of averaged null beams. At low averaging, the incoherent
reference keeps a modest S/N edge because it is intrinsically less noisy, even
though it self-subtracts a small amount of pulsar signal. As the null
reference is averaged over more beams and smoothed more heavily, its
variance drops and the ratio rises toward or above unity: null-division
preserves the pulsar signal, avoids self-subtraction, and cleans RFI
more completely. Survey pipelines that already downsample aggressively and can
average several null beams will therefore lie in or near the null-favouring
portion of the map; with few null beams or light smoothing, incoherent
subtraction remains slightly better in pure S/N despite its residual
self-subtraction.

\section{Simulations}
\label{sec:simulations}

The concept of beam flatfielding relies on the assumption that, for sources located outside the main lobe of a TAB, the residual phases at different antennas are effectively random and the power response converges toward the incoherent limit (Sect.~\ref{sec:beamforming}). The raw-voltage experiment in Sect.~\ref{sec:stations} probes this behaviour locally along a
single one-dimensional cut. To test the same assumptions in a realistic wide-field, multi-beam geometry, we use an existing survey TAB (TAB) pointing configuration as a worked example rather than designing a new layout from scratch.

Following \citet{san2019}, each LOFAR Tied-Array All-Sky Survey (LOTAAS)
pointing is divided into three sub-array pointings (SAPs), arranged on the
vertices of an equilateral triangle separated by 3\fdg82. Within each SAP COBALT generates 61 coherently summed survey TABs in a hexagonal pattern, plus 12 additional TABs pointed to known pulsars (if there are any in the field; if not, they are placed in a ring outside the central hexagonal grid), and
one incoherent beam. The TAB FWHM varies from 0\fdg41 to 0\fdg32 across the
LOTAAS band, while each SAP has an FWHM of 6\fdg1–4\fdg7. This is typical of
modern low-frequency pulsar and fast-transient surveys with aperture arrays:
a dense tiling of TABs across the station primary beam, augmented by
an incoherent beam for monitoring total power to do a more shallow, larger FoV survey of bright transients at the same time.

In the simulations a single SAP is reproduced with its 61 survey TABs and
12 pulsar TABs, for a total of 73 TABs. At a single representative
frequency the weighted coherence $R_w$ and normalized power response
$P_{\rm norm}$ are evaluated as a function of offset from a chosen reference
direction, taken here to be the sky position of PSR~J0250+5854.
The calculation is monochromatic and includes only geometric delays; in real
data it applies channel by channel, with frequency-dependent station gains
absorbed into the amplitude factors $A_i$.

\subsection{Mathematical setup}

Using the notation of Sect.~\ref{sec:beamforming}, we consider trial directions
$(\ell,m)$ across the SAP, with direction cosines defined relative to a chosen
boresight $(\ell_0,m_0)$. At each trial direction the tied-array voltage is
\begin{equation}
    b(\ell,m) = \sum_{i=1}^{N_a} A_i\,e^{\mathrm{i}\delta_i(\ell,m)},
    \label{eq:map_voltage}
\end{equation}
where $A_i$ is a real amplitude scale for sub-station $i$ and $\delta_i$ is the
residual phase relative to the boresight. The amplitudes are fixed by the
boresight response,
\begin{equation}
    A_i \equiv |V_i(\ell_0,m_0)|,
\end{equation}
with $V_i$ the complex station voltage in that direction, and the geometric
residual phases follow directly from the station coordinates $(u_i,v_i,w_i)$,
\begin{equation}
    \delta_i(\ell,m)
    = 2\pi \Big[ u_i(\ell_0-\ell) + v_i(m_0-m) + w_i(n_0-n) \Big],
    \label{eq:delta_geom}
\end{equation}
where $n=\sqrt{1-\ell^2-m^2}$ and $(\ell_0,m_0,n_0)$ are the boresight direction
cosines.

From Eq.~\ref{eq:map_voltage} we define two diagnostic maps,
\begin{align}
  R_w(\ell,m) &\equiv \frac{|b(\ell,m)|}{\sum_{i=1}^{N_a} A_i}, \\
  P_{\rm norm}(\ell,m) &\equiv
  \frac{|b(\ell,m)|^2}{\Big(\sum_{i=1}^{N_a} A_i\Big)^2}
  \;=\; R_w(\ell,m)^2.
\end{align}
Both are unity at boresight and fall as the source offset increases. The
incoherent floor is set by the effective number of contributing sub-stations,
\begin{equation}
  N_{\rm eff}(\ell,m) \equiv
  \frac{\Big(\sum_{i=1}^{N_a} A_i(\ell,m)\Big)^2}{\sum_{i=1}^{N_a} A_i^2(\ell,m)},
  \label{eq:Neff}
\end{equation}
so that far from boresight, where the residual phases decorrelate,
\begin{equation}
  P_{\rm norm}(\ell,m) \rightarrow \frac{1}{N_{\rm eff}(\ell,m)}.
\end{equation}

\subsection{Diagnostics and interpretation}

Three diagnostics are used to illustrate the transition from coherent to
incoherent response across the SAP.

\begin{figure*}
  \centering
  \includegraphics[width=\linewidth]{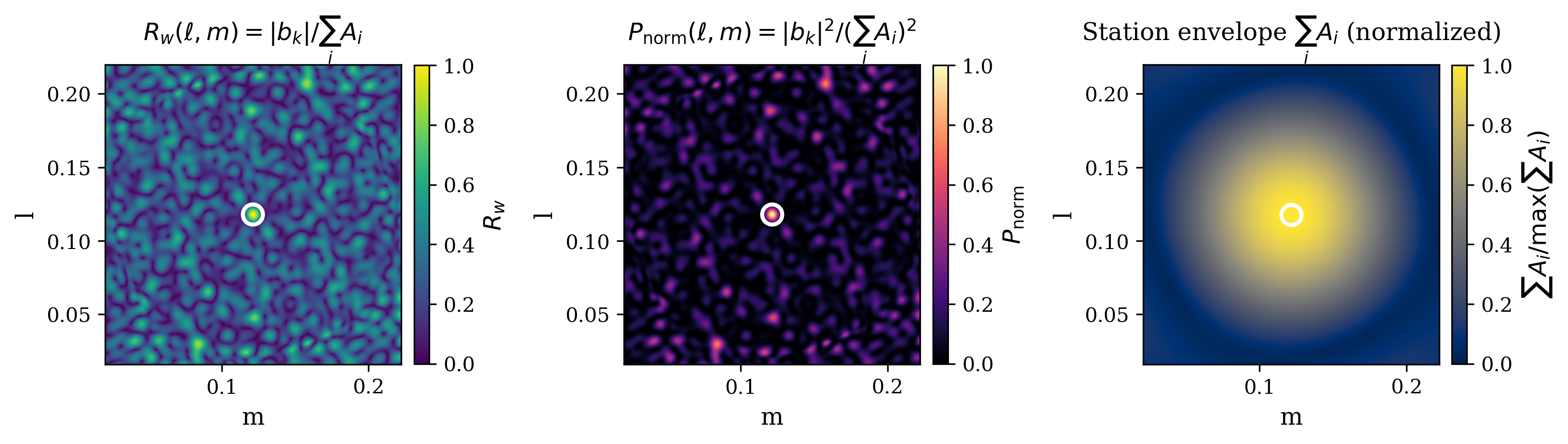}
  \caption{Boresight-centred diagnostics for one LOTAAS-like SAP:
  left, weighted coherence $R_w(\ell,m)$; middle, normalized power
  $P_{\rm norm}(\ell,m)$; right, primary-beam envelope
  $\sum_i A_i(\ell,m)$ (normalized). The white circle marks the boresight.}
  \label{fig:coh_maps}
\end{figure*}

\paragraph{(i) Boresight-centred maps.}
Figure~\ref{fig:coh_maps} shows maps of $R_w(\ell,m)$,
$P_{\rm norm}(\ell,m)$, and the envelope $\sum_i A_i(\ell,m)$ on a continuous
grid centred at the boresight. By construction, all maps peak at
$(\ell_0,m_0)$.This figure demonstrates that the coherent array gain is confined to a narrow region around the boresight. Beyond this region, $R_w$ and $P_{\rm norm}$ flatten out at the incoherent floor, and the response is dominated entirely by the station's primary-beam envelope (right panel). This confirms that global sky variations will appear - on average - identical in these regions regardless of the specific beam steering direction.

\begin{figure*}
  \centering
  \includegraphics[width=\linewidth]{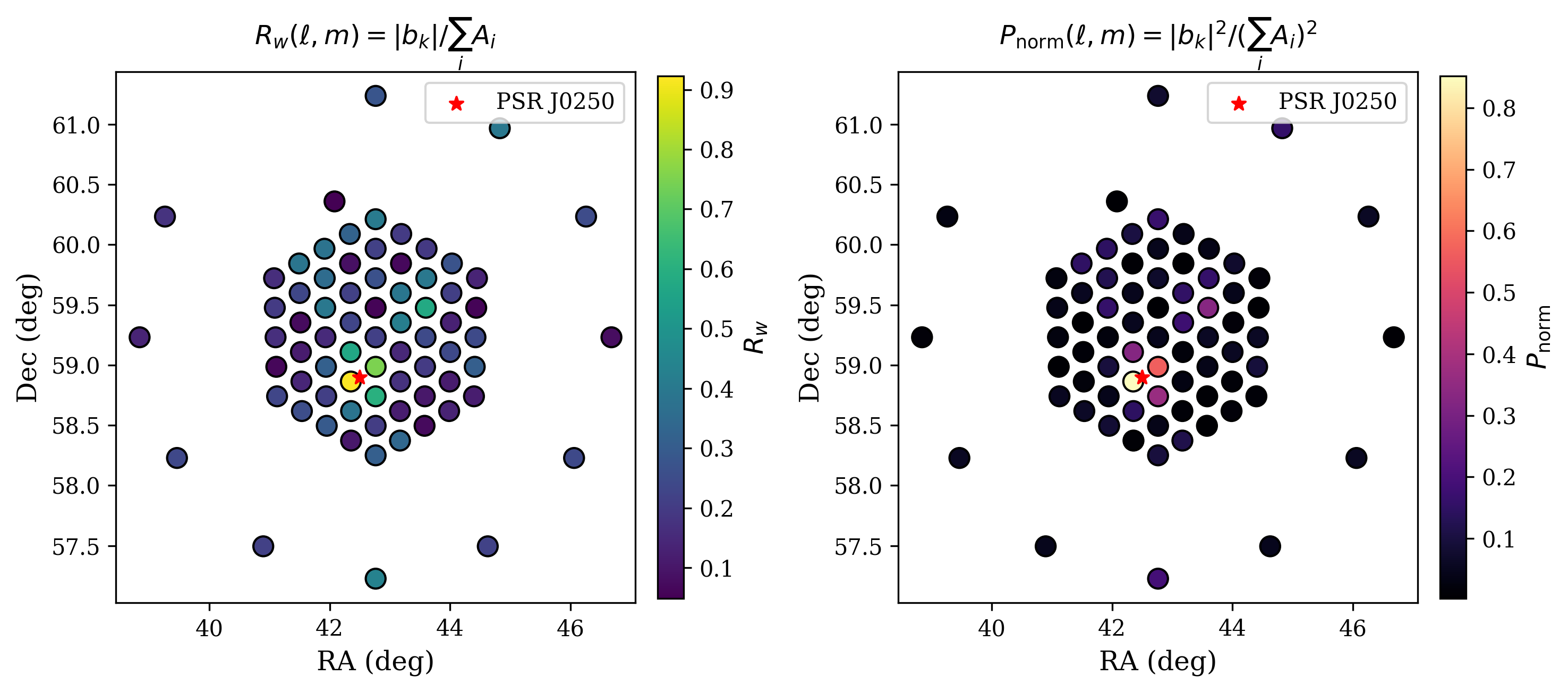}
  \caption{TAB layout within one SAP. Each circle marks a TAB at its sky
  position, coloured by weighted coherence
  $R_w=|b_k|/\sum_i A_i$ (left) and normalized power
  $P_{\rm norm}=|b_k|^2/(\sum_i A_i)^2$ (right). The red star marks the sky
  position of PSR~J0250+5854, used here as the reference direction. Outer
  beams converge toward the incoherent limit.}
  \label{fig:beam_scatter}
\end{figure*}

\paragraph{(ii) Per-beam responses.}
Figure~\ref{fig:beam_scatter} shows the 73 TAB positions within one SAP,
coloured by the simulated $R_w$ (left) and $P_{\rm norm}$ (right). The red
star marks the sky position of PSR~J0250+5854, which is also the adopted
reference direction in this simulation. Beams close to that position retain
high coherence and near-unity normalized power, while beams further out cluster
around the incoherent level set by $1/N_{\rm eff}$.

\begin{figure}
  \centering
  \includegraphics[width=\linewidth]{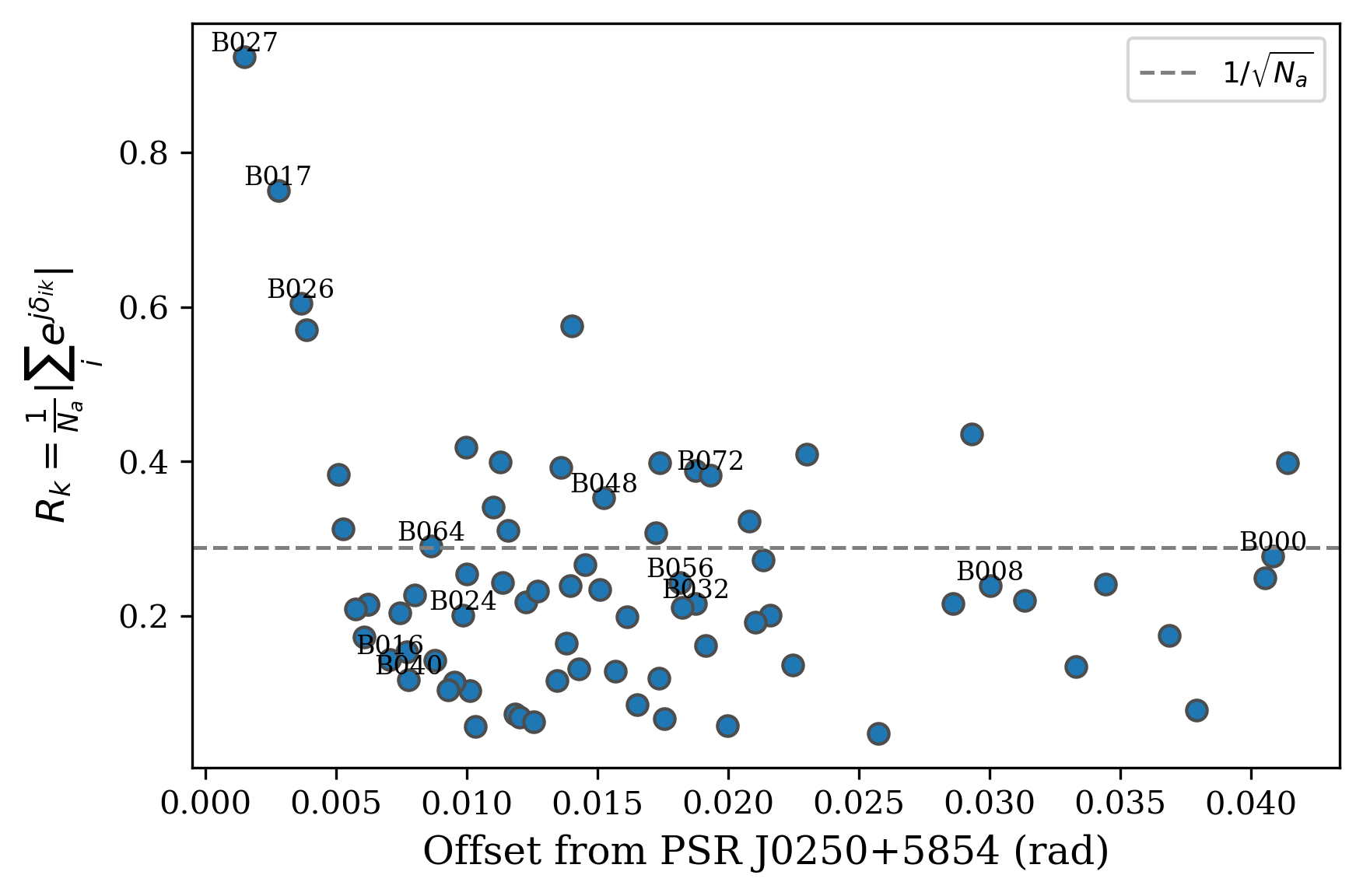}
  \caption{Phase concentration $R_k$ as a function of angular offset from the
  sky position of PSR~J0250+5854, used here as the reference direction. Beams
  near that position show high concentration; outer beams have low $R_k$,
  consistent with random residual phases.}
  \label{fig:phase_concentration}
\end{figure}

\paragraph{(iii) Phase concentration.}
Figure~\ref{fig:phase_concentration} shows the phase concentration
\begin{equation}
    R_k = \left|\frac{1}{N_a}\sum_i e^{j\delta_{ik}}\right|
\end{equation}
for each TAB $k$ as a function of angular offset from the adopted reference
direction, here taken to be the sky position of PSR~J0250+5854. Beams near
that position retain high concentration ($R_k\simeq1$), while outer beams have
low $R_k$, consistent with their residual phases being effectively random.

Together these diagnostics confirm, on an existing wide-field survey geometry,
the assumption used in Sect.~\ref{sec:flatfielding}: for sources outside the
main lobe the residual station phases are effectively random and the tied-array
intensity follows the incoherent limit set by $N_{\rm eff}$. The inputs to
this calculation are generic (station layout, TAB tiling, coherent weights), so
the conclusions transfer directly to other multi-beam tied-array surveys with
similar configurations. This justifies treating off-target astrophysical sources
and broad-band RFI as nearly uniform contributors across beams and underpins
the beam flatfielding approach used here.

\section{Verification on real survey data: LOTAAS case study of PSR~J0250+5854}
\label{sec:realdata}

The simulations in Sect.~\ref{sec:simulations} demonstrated that, for sources 
outside the main lobe, the tied–array power converges toward the incoherent limit.  
We now confirm this behaviour directly in observational data and assess its impact 
on single–pulse searches and folded detections in a standard survey pipeline.
PSR~J0250+5854 is a particularly stringent test case for this purpose. It is an
ultra-slow, rotation-powered pulsar with $P=23.53$~s and a pulse width of order
 $\sim 300$~ms, so low-frequency red noise in the fluctuation spectrum and
slow baseline variations in the time series and dynamic spectrum fall directly
in the parameter range most relevant to its detection. In the original LOTAAS
processing it was found only because it was bright enough to be recovered
through a harmonic response \citep{tan2018}. This makes it a natural example
for showing that beam flatfielding is useful not only for reducing false
positives within the parameter space already searched for pulsars and fast
transients, but also for extending that search toward longer periods and broader
pulses where red noise and baseline drifts otherwise dominate.

\subsection{Measured beam response pattern}

\begin{figure}
  \centering
  \includegraphics[width=\linewidth]{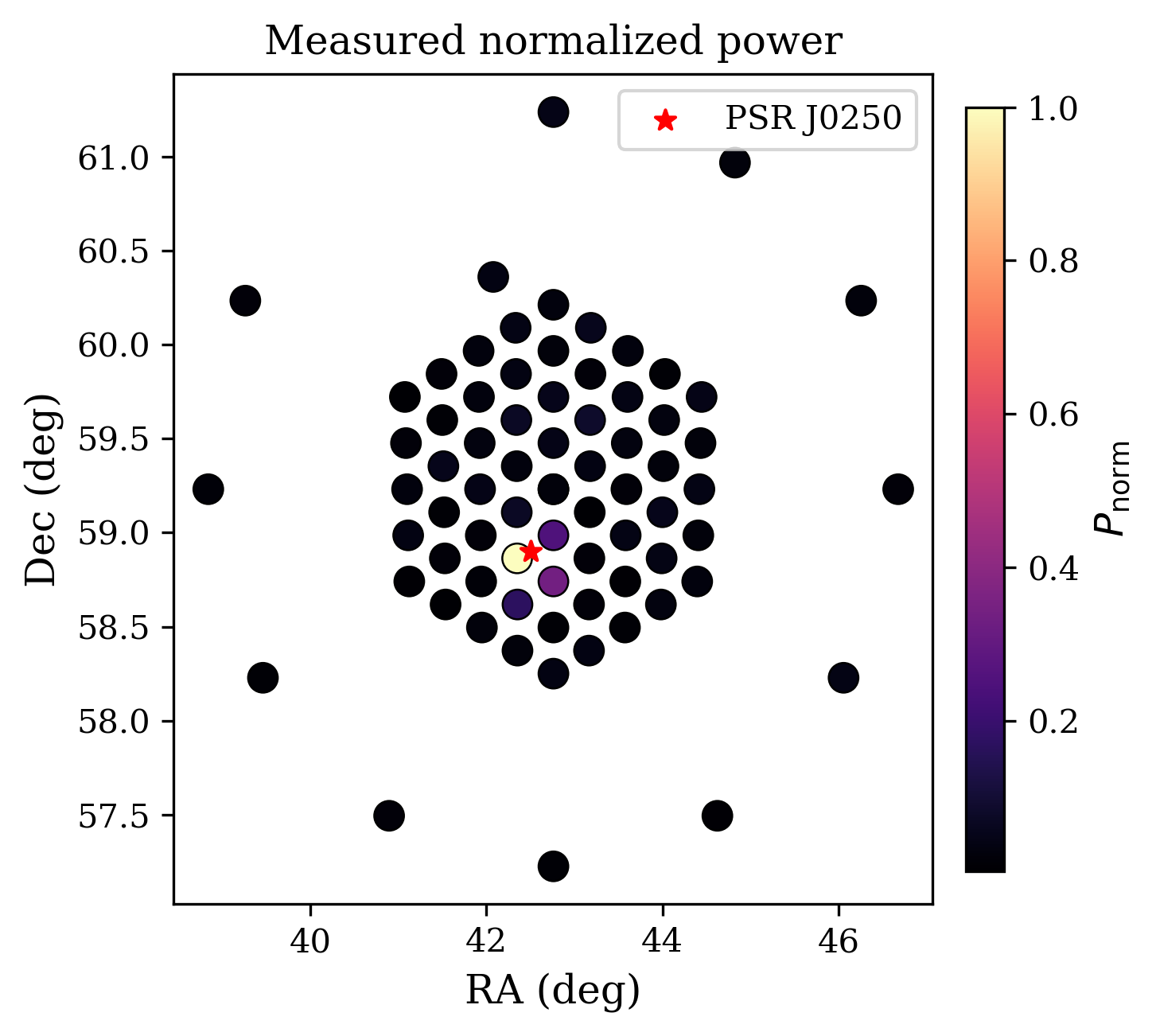}
  \caption{Measured normalized power across all TABs in the LOTAAS 
  sub–array pointing containing PSR~J0250+5854. Each circle represents one TAB, 
  coloured by its folded S/N (integrated for the full 1-hr observation), and normalized to the maximum. The red star marks the 
  pulsar position. The observed beam pattern closely matches the simulated response 
  in Fig.~\ref{fig:beam_scatter}.}
  \label{fig:beam_layout_real}
\end{figure}

Figure~\ref{fig:beam_layout_real} shows the measured normalized power across all
TABs within the LOTAAS SAP containing PSR~J0250+5854.  
For each beam, we compute the folded S/N of the pulsar at its known period
($P=23.53$~s) and dispersion measure (DM = 45.2~pc\,cm$^{-3}$), and normalize the
values to the maximum over all beams.  
The resulting map reproduces the structure predicted by the simulations
(Fig.~\ref{fig:beam_scatter}): central beams occupy the high-response core,
while the outer beams fall smoothly toward the incoherent floor.  
This empirical agreement confirms that the random–phase assumption underpinning
beam flatfielding holds in real observations, providing a direct observational validation 
of the simulated behaviour.

\subsection{Single–pulse diagnostics}

Having confirmed that the beam response behaves as expected, we next examine the
effect of beam flatfielding on single–pulse search results for the same observation.
The dataset was processed twice using an otherwise identical pipeline:
(1) starting from the original tied–array beam, and (2) starting from the same
beam after applying the beam flatfielding defined in 
Eq.~(\ref{eq:ff_main_compact}). In both cases the normalization is applied to the
filterbank data before running \textsc{PRESTO}.

The single–pulse search was carried out using the standard \textsc{PRESTO} pipeline
\citep{ransom2002}\footnote{\url{https://www.cv.nrao.edu/~sransom/presto/}}, beginning with RFI excision via \texttt{rfifind} and subsequent
dedispersion into trial DMs using \texttt{prepsubband}. Each dedispersed time series
was convolved with a range of boxcar filters using \texttt{single\_pulse\_search.py}
to match pulse widths from a few to several hundred milliseconds, and all detections
above a threshold of $\mathrm{S/N}=5$ were recorded. The resulting candidate lists were
then merged and analyzed to produce diagnostic plots of S/N, DM, and time
distributions, enabling a direct comparison between the original and beam flatfielded data.

\begin{figure*}
  \centering
  \includegraphics[width=\textwidth]{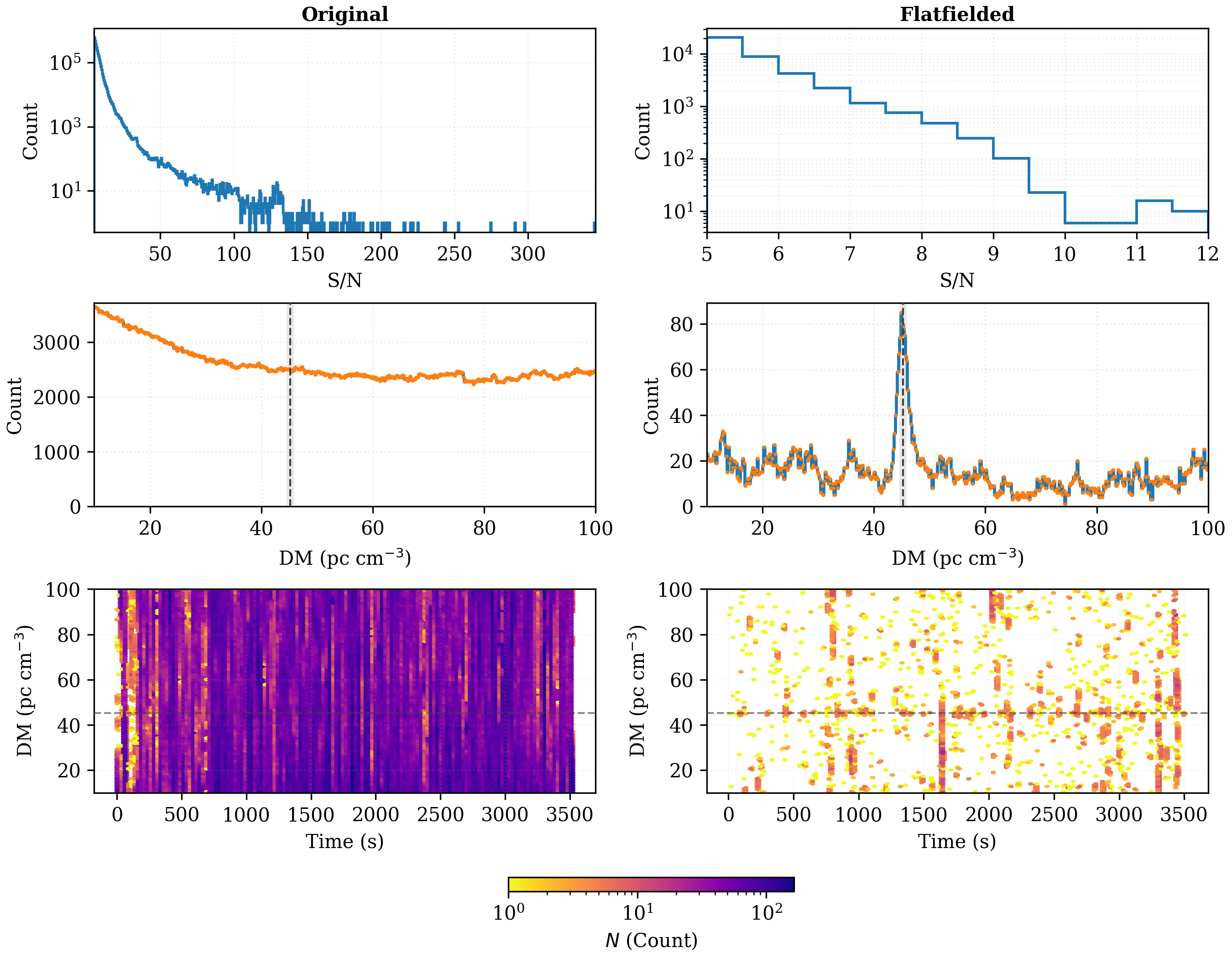}
  \caption{Single–pulse overview for PSR~J0250+5854. \emph{Left:} original TAB; \emph{right:}
  the same TAB after beam flatfielding. Top: S/N histograms (log–scaled, threshold at S/N=5).
  Middle: number of candidates vs.\ DM (10–100~pc\,cm$^{-3}$); dashed line indicates the expected
  DM. Bottom: time vs.\ DM density maps (shared colourbar indicates $\log_{10}N$). Beam flatfielding
  suppresses off–DM triggers by more than two orders of magnitude and sharpens the pulsar peak to a
  narrow feature centred at the true DM.}
  \label{fig:overview_j0250}
\end{figure*}

Figure~\ref{fig:overview_j0250} summarises the outcome. Each column corresponds 
to one processing mode (left: original; right: flatfielded). 
The top row shows S/N histograms, the middle row candidate counts versus DM, 
and the bottom row the time–DM density of detections.  

Within 10–100~pc\,cm$^{-3}$, the total number of detected events decreases from
$1.16\times10^{6}$ in the original data to $6.9\times10^{3}$ after beam flatfielding.
Off–DM events (i.e.\ $|{\rm DM}-45.2|>0.5$~pc\,cm$^{-3}$) drop from $1.15\times10^{6}$ to
$6.5\times10^{3}$, corresponding to a factor of $\sim\!1.8\times10^{2}$ reduction in false
positives. This reflects a strong suppression of spurious broad–band fluctuations, while
leaving the on–DM signal from PSR~J0250+5854 clearly visible. In terms of the noise
statistics discussed in Sect.~\ref{sec:flatfielding} and Appendix~\ref{app:noise_stats},
the beam flatfielding step removes large–scale multiplicative structure and narrows the noise distribution at fixed S/N threshold.

\subsection{Red noise reduction}
\label{subsec:rednoise}

\begin{figure}[t]
  \centering
  \includegraphics[width=\linewidth]{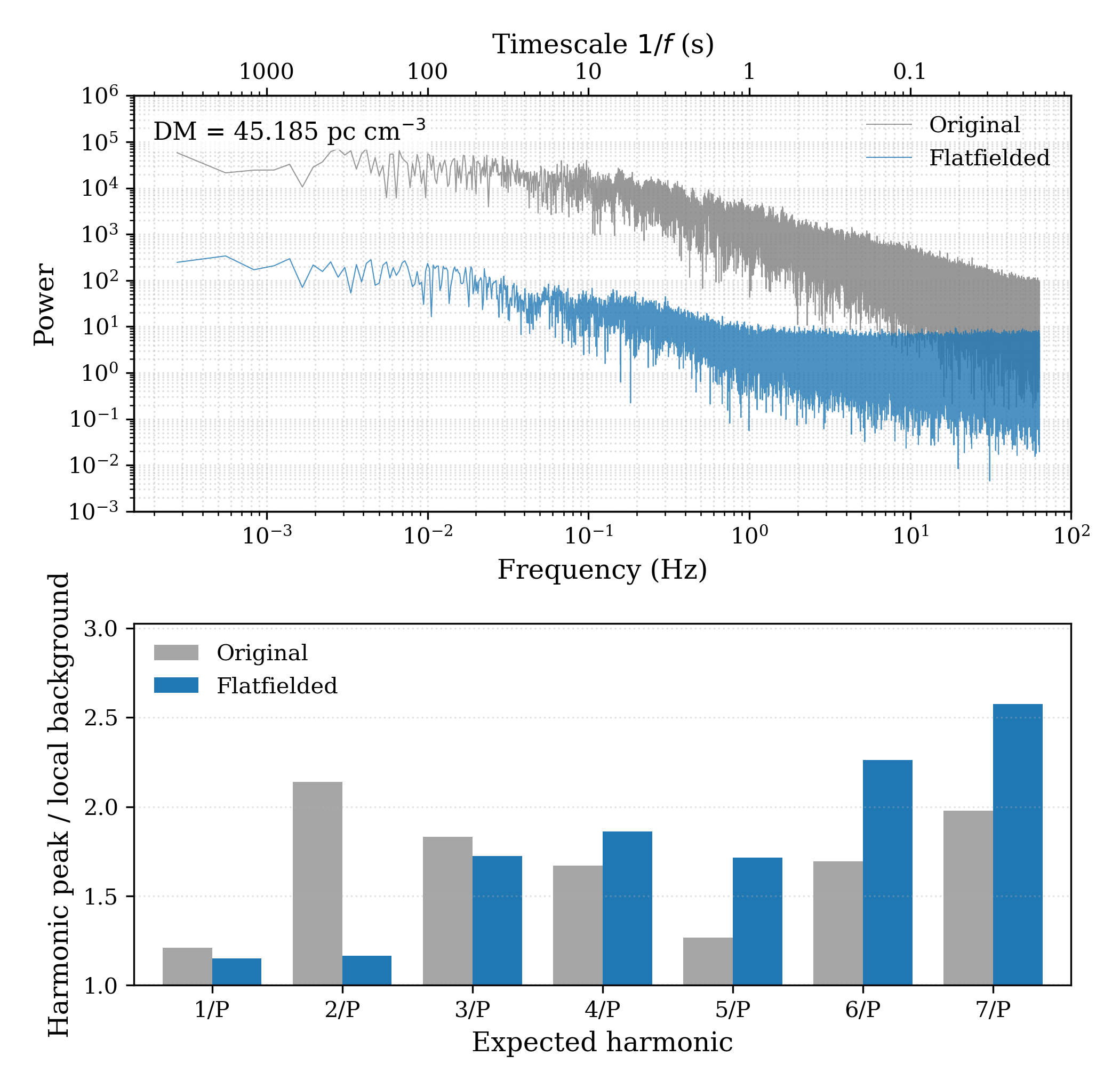}
  \caption{Red-noise suppression and harmonic detectability in the data containing PSR~J0250+5854. Top: Fourier spectrum of the time series dedispersed to the pulsar DM, with the top axis giving the equivalent fluctuation timescale, $1/f$, and the bottom axis giving the fluctuation frequency. Beam flatfielding suppresses the low-frequency continuum associated with slow baseline drifts and broad-band RFI, so the spectrum remains closer to its flatter high-frequency baseline out to lower frequencies before turning upward. Bottom: the first seven expected pulsar harmonics, where each bar gives the ratio of the peak amplitude near $n/P$ to the local sideband background around that harmonic. The lowest harmonics remain the most contaminated by residual low-frequency structure, but the higher harmonics 4/P--7/P, corresponding to fluctuation timescales of 5.9--3.4~s, are all more distinct after flatfielding.}
  \label{fig:rednoise}
\end{figure}

A key advantage of beam flatfielding is the suppression of correlated low-frequency variability, often referred to as ``red noise''. This noise arises from gain fluctuations, atmospheric effects, and broad-band RFI that are common across the array. Figure~\ref{fig:rednoise} therefore focuses directly on the Fourier spectrum of the time series dedispersed to the pulsar DM. The upper panel shows the full spectrum, with the top horizontal axis giving the corresponding fluctuation timescale, $1/f$, and the bottom horizontal axis giving the fluctuation frequency. In this broadband view the beam flatfielded spectrum shows a strong reduction in the low-frequency continuum, corresponding to timescales from roughly 1 to 1000~s. Put another way, as one moves leftward to lower fluctuation frequencies, the original spectrum turns upward much earlier, whereas the flatfielded spectrum stays close to its flatter high-frequency baseline over a wider range. The red-noise upturn is therefore pushed to lower frequencies, or equivalently to longer timescales, after flatfielding. That panel is too broad, however, to judge the detectability of the pulsar itself directly. The lower panel therefore examines the specific Fourier frequencies at which the pulsar is expected to appear, isolating the first seven harmonics and comparing each one through the ratio of the local peak amplitude to the surrounding sideband background. This makes the trade-off clearer: the lowest harmonics, which correspond to the longest timescales and remain embedded in the messiest part of the red-noise continuum, can still appear somewhat stronger in the original spectrum. At higher harmonics, however, the local continuum has been more strongly suppressed, and the contrast improves in the flatfielded data. In particular, the 4/P--7/P harmonics, corresponding to fluctuation timescales of about 5.9--3.4~s, are all stronger after flatfielding.

\subsection{Consistency check: periodic folding}

\begin{figure*}
  \centering
  \includegraphics[width=\textwidth]{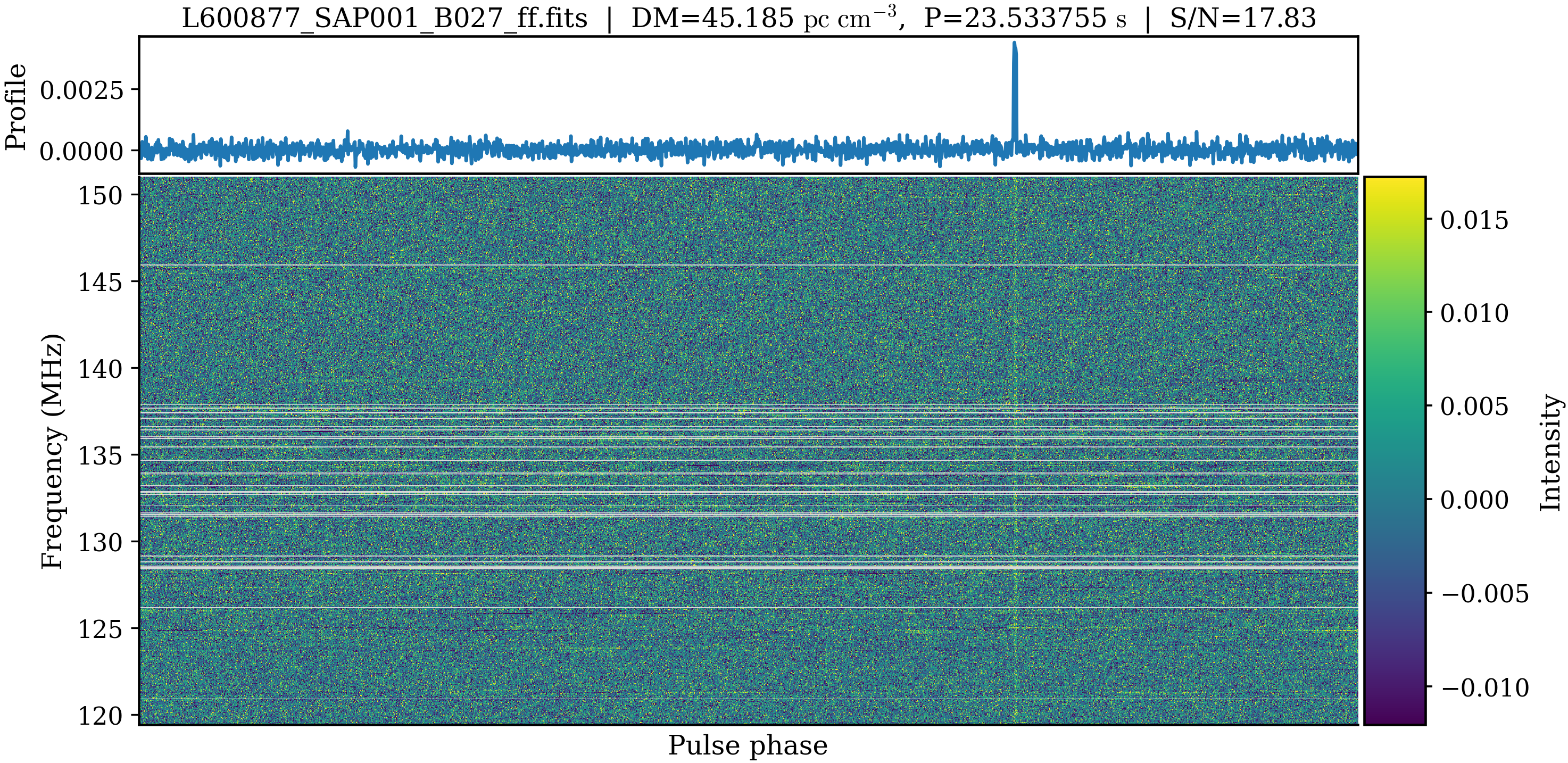}\hfill
  \includegraphics[width=\textwidth]{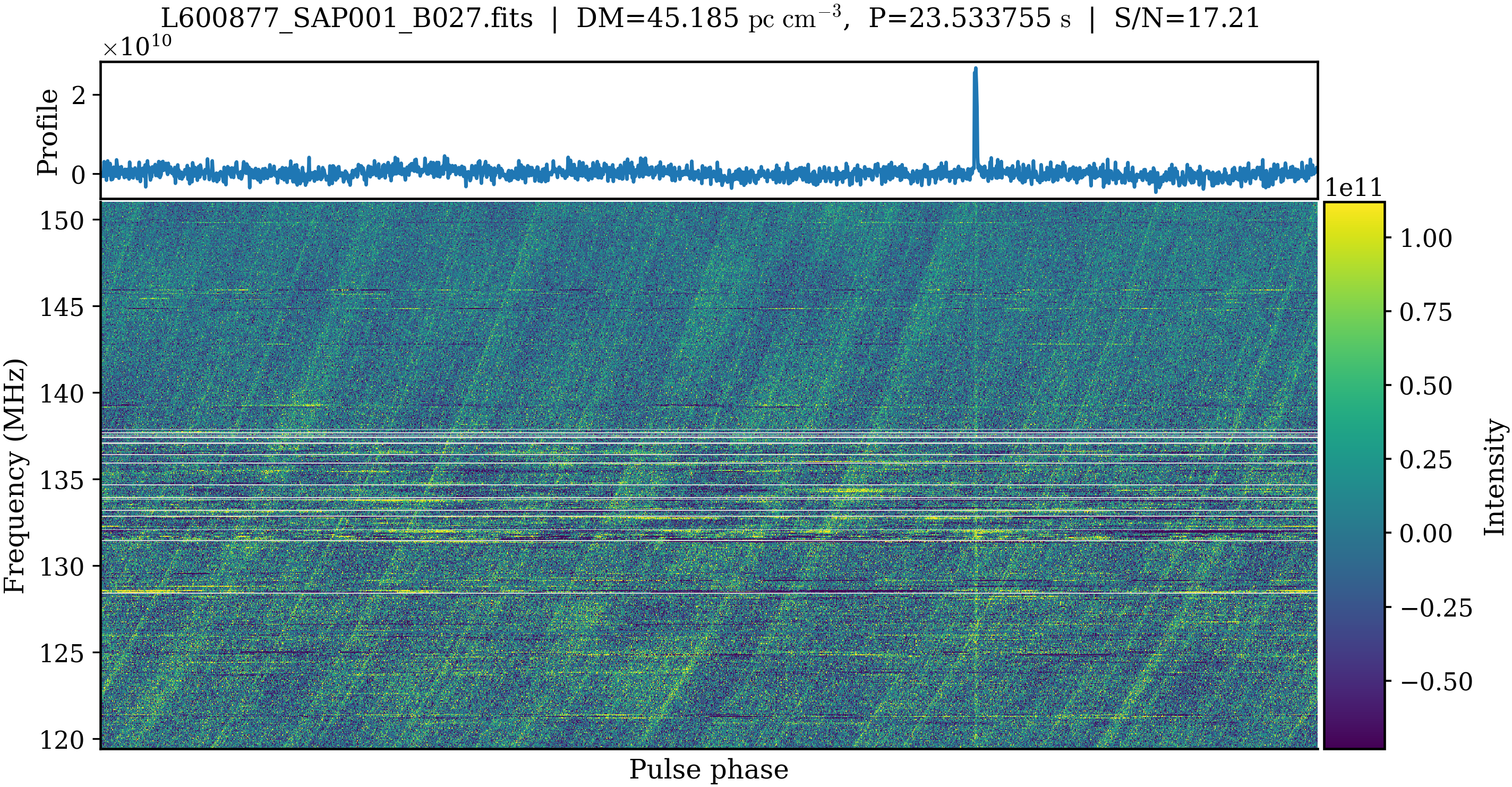}
  \caption{Folded dynamic spectra and frequency–summed profiles for PSR~J0250+5854.
  \emph{Top:} flatfielded data; \emph{Bottom:} original.
  Both are dedispersed to the known DM and folded using a rotational timing ephemeris.
  Beam flatfielding regularises the bandpass and stabilizes the baseline across the band,
  producing a cleaner frequency–summed profile at comparable peak S/N for the pulsar.}
  \label{fig:fold_comp}
\end{figure*}

Beam flatfielding is motivated not only by improved single–pulse sensitivity and
reduced candidate false positives, but also by the suppression of low-frequency
red noise shown in Fig.~\ref{fig:rednoise}. That figure also shows why this
matters for slow periodicity searches: the lowest harmonics remain the most
contaminated, but the higher harmonics that fall in a cleaner part of the
spectrum become more distinct after flatfielding. This should improve
periodicity searches for slow pulsars in both FFT-based and fast-folding
pipelines. It is therefore important to confirm that the normalization preserves the underlying
astrophysical signal. To test this, we fold the same TAB data at the known
ephemeris of PSR~J0250+5854 and compare the resulting dynamic spectra and
profiles before and after beam flatfielding (Fig.~\ref{fig:fold_comp}).

The flatfielded data exhibit a markedly steadier baseline across the band and a cleaner
frequency–summed profile, consistent with the removal of large–scale multiplicative gain
variations discussed in Sect.~\ref{sec:flatfielding}. The peak S/N remains comparable between the
two, indicating that the normalization does not significantly alter the pulse amplitude, while the off–pulse
variance is reduced, yielding a flatter baseline. After RFI excision, the number of usable
frequency channels increases from 586 of 648 (90.4\%) to 618 of 648 (95.4\%), corresponding to an
improvement of 32 channels (+4.9 percentage points) and a factor 2.1 reduction in flagged channels.
Together, these results confirm that beam flatfielding preserves the intrinsic pulse morphology while
producing data of higher statistical quality.

\section{Discussion and conclusions}
\label{sec:discussion}

The goal of this work was to quantify and exploit the spatial information in
multi-beam tied-array observations for time-domain searches. We showed that for
sources outside the main lobe of a TAB, the residual phases
between stations are effectively random and the array response converges toward the
incoherent limit. In this regime, off-target point sources and broad-band,
near-field RFI contribute on average nearly uniform power to every simultaneously formed
beam, modulated only by the slowly varying station primary beams and bandpasses.
This motivates beam flatfielding, in which the dynamic spectrum of each beam is
divided by a reference constructed from the other beams.

Using a simple statistical model for tied-array and incoherent beams
(Sect.~\ref{sec:flatfielding} and Appendix~\ref{app:noise_stats}), we derived
the expected impact of such a normalization on the noise properties and the
effective S/N ratio. The key point is that TABs have
low-degree-of-freedom $\chi^2$ noise, whereas an incoherent sum, or a suitably
smoothed average over many beams, approaches a high-degree-of-freedom
$\chi^2$ distribution with much smaller fractional fluctuations. As a result, once
the reference has been averaged over a modest number of beams and over the
time-frequency scales already used in standard search pipelines, the additional
variance introduced by dividing by a noisy reference is negligible compared to the
intrinsic noise of the target TAB. The dominant effect of
beam flatfielding is then the removal of large-scale multiplicative structure, which
shrinks non-Gaussian wings in the noise distribution at fixed S/N threshold.

We tested these expectations in two complementary regimes. First, we used raw
voltages of PSR~B0329+54 to form tied-array, incoherent, and deliberately placed
null beams offline (Sect.~\ref{sec:stations}). This allowed us to construct
null-beam references that strongly suppress the on-axis pulsar while sharing the
same bandpass, gain drifts, and broad-band RFI as the target beam. We compared
null division and incoherent subtraction. In the configuration shown in
Fig.~\ref{fig:dyn_spectra}, null division gives a slightly higher robust S/N
than incoherent subtraction ($9.1$ versus $8.8$) while avoiding the $\sim10$-$13\%$
self-subtraction of source power introduced by the incoherent reference. The
broader grid search in Fig.~\ref{fig:regime_map} shows that this balance depends
on the stability of the null-beam reference: with only a few null beams or light
smoothing, incoherent subtraction can retain a modest S/N advantage, but as the
null reference is averaged over more beams and time samples, null division reaches
and then exceeds parity. This reflects the higher thermal noise of individual
null beams; aggressive smoothing is needed before division fully avoids a noise
penalty.

Second, we simulated a full LOTAAS-style sub-array pointing using the actual
TAB geometry (Sect.~\ref{sec:simulations}). These simulations
confirm that beams near the pointing centre remain fully coherent, while outer
beams rapidly converge to the incoherent limit once the station primary-beam
envelope is accounted for (Figs.~\ref{fig:coh_maps}-\ref{fig:phase_concentration}).
The per-beam coherence and normalized power cluster into two regimes: a small
set of highly coherent beams on and around the source and a much larger set of outer beams whose
responses are effectively incoherent. This is the configuration in which
beam flatfielding works well: a few beams carry most of the on-axis
signal, while the majority provide a stable estimate of the common multiplicative
structure.

We then applied beam flatfielding to real LOTAAS data of the ultra-slow
pulsar PSR~J0250+5854 (Sect.~\ref{sec:realdata}). The measured per-beam folded
S/N pattern (Fig.~\ref{fig:beam_layout_real}) reproduces the simulated response
across the tied-array field, directly validating the random-phase assumption
for off-target sources in a survey configuration. Running an otherwise identical
\textsc{PRESTO} single-pulse search on the original and beam flatfielded data shows
the practical impact on a real survey pipeline: within 10-100~pc\,cm$^{-3}$,
the number of off-DM single-pulse triggers is reduced by a factor of
$\sim 1.8\times10^2$, while the pulsar remains clearly detected at the correct
DM (Fig.~\ref{fig:overview_j0250}). Periodic folding confirms that the pulse
amplitude and morphology are preserved and that the main effect of beam flatfielding
is to stabilize the baseline and recover additional usable channels
(Fig.~\ref{fig:fold_comp}). Spectral analysis (Sect.~\ref{subsec:rednoise}) confirms that this stability extends to minute-long timescales, effectively suppressing low-frequency red noise. This capability is particularly valuable for detecting transients with durations of seconds or longer. While millisecond-scale transients are spectrally distinct from slow instrumental gain drifts and can often be recovered with standard temporal high-pass filters, slower transients occupy the same spectral window as these instrumental variations. A useful way to view this is that once the dispersive delay across the band becomes small compared to the pulse width, dedispersion no longer provides an effective time-frequency filter for separating astrophysical signals from broadband RFI. By removing these drifts spatially rather than temporally, beam flatfielding cleans the search window to find slow events like those from PSR~J0250+5854 that would otherwise be buried in red noise, without the signal loss associated with aggressive temporal filtering. Together, these results demonstrate that
beam flatfielding can reduce candidate false positives by nearly two
orders of magnitude without degrading astrophysical signals. In that sense,
the PSR~J0250+5854 case study illustrates not only a reduction in false
positives within the conventional search space, but also an expansion of the
accessible search space toward longer periods and broader pulses that are
otherwise strongly limited by red noise and baseline instabilities.

The method has several practical advantages. It operates entirely in beam space
on detected Stokes~$I$ filterbank data and requires only simple statistics
(mean or robust mean) across beams and time-frequency bins. There is no need to
access visibilities or to modify upstream correlator and beamformer hardware.
The computational cost is negligible compared to dedispersion and matched
filtering, and in practice it can usually be applied before or after most
existing RFI excision steps, so it can be inserted into current pipelines with
minimal changes. Beam
flatfielding is also complementary to post-correlation beamforming and other
visibility-domain schemes that reduce red noise and RFI at the correlation
stage; it can be applied regardless of how the TABs were formed, as
long as multiple simultaneous beams are available.

There are, however, clear limitations and failure modes that need to be
acknowledged. The approach assumes that the multiplicative structure to be
removed is common to most beams. Strong, spatially localised RFI that affects
only a subset of stations or beams will not be removed and can even be
emphasised if it enters the reference. Likewise, beams with dropped packets,
missing samples, or other instrumental or numerical artefacts that do not
reflect received sky power can bias the reference if they are not flagged in
advance. Similarly, very bright sources that are present in multiple
neighbouring beams can contaminate the reference unless nearby beams are
excluded or downweighted. In practice, these issues can be mitigated by using
robust estimators (e.g., trimmed or median averages across beams), by excluding
beams that fail simple data-quality checks, by excluding a small neighbourhood
around each target beam from the reference, and by combining beam flatfielding
with standard per-beam RFI masking.
The method is also less effective for arrays with only a handful of beams or for
survey modes in which beams sparsely tile the primary beam and do not provide a
dense sampling of the shared multiplicative structure.

Looking ahead, beam flatfielding is directly applicable to current and future
multi-beam time-domain surveys that already form tens to hundreds of beams per
pointing. For LOFAR, we are currently implementing this technique in the full
reprocessing of the LOTAAS survey to improve sensitivity to long-period pulsars
and transients. Similarly, the method is being adapted for the CHIME/Slow
survey \citep{mate_inprep}, which targets slow transients up to
second-duration timescales. In that regime, slow baseline structure in the beam
time series can directly mimic or bury broad events, so beam flatfielding
should be especially valuable not only for stabilising the bandpass but also
for flattening the time-series baseline before candidate generation. For
MeerKAT, SKA-Low, and related phased-array systems, the same
principles apply: as long as the beam layout yields a large population of beams
that are incoherent to a given source, those beams can be used to stabilize the
bandpass and suppress broad-band RFI
in the beams of interest. Future work should explore more systematic survey
designs that explicitly reserve null beams or low-coherence regions of the
sub-station beam pattern for beam flatfielding, and should quantify the impact on
fast-transient detection statistics in more realistic RFI environments and at
higher time and frequency resolution. It will also be useful to combine
beam flatfielding with downstream machine-learning classifiers, where a
cleaner, more Gaussian noise distribution should translate directly into better
control over false-alarm rates at fixed sensitivity.
An important extension is to test whether the same beam flatfielding normalization
works for the full set of Stokes parameters. Polarisation leakage and Faraday
rotation will imprint different common structures on $Q/U/V$, so it will be
valuable to quantify how well a beam-averaged reference stabilizes those
channels and whether robust or polarisation-aware references are needed.
Related work includes more robust reference constructions (e.g., trimmed or
median means across beams), adaptive beam selection that downweights beams near
bright sources or localised RFI, and tests at higher time/frequency resolution
where denominator stability is most demanding.

In summary, we have shown that the spatial redundancy inherent in modern
multi-beam observations can be used as an effective, low-cost
lever to reduce red noise and RFI in time-domain searches. Beam
flatfielding exploits the fact that most beams see the same instrument and sky
at any given time and frequency. By using those beams as a reference, we can
regularise the dynamic spectra of the beams of interest, preserve astrophysical
signals, and substantially reduce the number of spurious candidates that have to
be sifted and classified by more computationally expensive stages in the pipeline.

\section*{Data availability}
The scripts, reduced data products, and frozen figure files required to
reproduce the figures and main results of this work are available on Zenodo at
\url{https://doi.org/10.5281/zenodo.18980491}. The archive contains the Python
source code, the reduced beamforming products used for the beamforming and
flatfielding analyses, the derived beam-S/N table used for the real-data beam
layout figure, the reduced FITS products used for the red-noise and folded
profile figures, and the PRESTO single-pulse candidate files used for the
overview comparison. The raw LOFAR data underlying this study were obtained
under project codes LT5\_004 and LC7\_018 and are available subject to the
LOFAR-ERIC data-access policies.

\begin{acknowledgements}

CGB acknowledges useful discussions with Albert-Jan Boonstra. We thank Tia Nolan and Jeff Huang for insightful discussions on using beam flatfielding for CHIME data.

The AstroFlash research group at McGill University, University of Amsterdam, ASTRON, and JIVE is supported by: a Canada Excellence Research Chair in Transient Astrophysics (CERC-2022-00009); an Advanced Grant from the European Research Council (ERC) under the European Union's Horizon 2020 research and innovation programme (`EuroFlash'; Grant agreement No. 101098079); an NWO-Vici grant (`AstroFlash'; VI.C.192.045); an ERC Starting Grant (`EnviroFlash'; Grant agreement No. 101223057); and an NWO-Veni grant (VI.Veni.222.295).

This paper is based (in part) on data obtained with the LOFAR telescope (LOFAR-ERIC) under project codes LT5\_004 and LC7\_018. LOFAR \citep{haa2013} is the Low Frequency Array designed and constructed by ASTRON. It has observing, data processing, and data storage facilities in several countries, that are owned by various parties (each with their own funding sources), and that are collectively operated by the LOFAR European Research Infrastructure Consortium (LOFAR-ERIC) under a joint scientific policy. The LOFAR-ERIC resources have benefited from the following recent major funding sources: CNRS-INSU, Observatoire de Paris and Université d'Orléans, France; Istituto Nazionale di Astrofisica (INAF), Italy; BMBF, MIWF-NRW, MPG, Germany; Science Foundation Ireland (SFI), Department of Business, Enterprise and Innovation (DBEI), Ireland; NWO, The Netherlands; The Science and Technology Facilities Council, UK; Ministry of Science and Higher Education, Poland.

This work used the Dutch national e-infrastructure with the support of the SURF Cooperative using grant no. EINF-14308.
\end{acknowledgements}

\bibliographystyle{aa}
\bibliography{references}

\begin{appendix}

\section{Expected array gain as a function of residual phase dispersion}
\label{app:array_gain}

This appendix details the derivation of the expected tied-array power
$\mathbb{E}[I_k]$ as a function of the rms residual phase $\sigma_\delta$.

Starting from Eq.~\ref{eq: beam_power} in Sect.~\ref{sec:beamforming}, the beam
power in beam $k$ is
\begin{equation}
    I_k = |A|^2 \left|\sum_{i=1}^{N_a} \exp(\mathrm{i}\delta_{ik})\right|^2,
\end{equation}
where $\delta_{ik} = \psi_i - \phi_{ik}$ is the residual phase at station $i$ in
beam $k$. Expanding the squared magnitude,
\begin{align}
    I_k
    &= |A|^2 \left( \sum_{i=1}^{N_a} \exp(\mathrm{i}\delta_{ik}) \right)
              \left( \sum_{m=1}^{N_a} \exp(-\mathrm{i}\delta_{mk}) \right) \nonumber\\
    &= |A|^2 \sum_{i=1}^{N_a} \sum_{m=1}^{N_a}
         \exp\left[\mathrm{i}(\delta_{ik}-\delta_{mk})\right].
\end{align}

It is useful to view the terms in the double sum as entries of a Hermitian
matrix
\begin{equation}
\mathbf{M}_k =
\begin{bmatrix}
1 & e^{\mathrm{i}(\delta_{1k}-\delta_{2k})} & \dots & e^{\mathrm{i}(\delta_{1k}-\delta_{N_ak})} \\
e^{\mathrm{i}(\delta_{2k}-\delta_{1k})} & 1 & \dots & e^{\mathrm{i}(\delta_{2k}-\delta_{N_ak})} \\
\vdots & \vdots & \ddots & \vdots \\
e^{\mathrm{i}(\delta_{N_ak}-\delta_{1k})} &
e^{\mathrm{i}(\delta_{N_ak}-\delta_{2k})} & \dots & 1
\end{bmatrix},
\end{equation}
so that $I_k/|A|^2$ is simply the sum of all elements of $\mathbf{M}_k$. The
diagonal elements equal unity and the off-diagonal elements encode the phase
differences between stations. Splitting the sum into diagonal and off-diagonal
contributions gives
\begin{equation}
    I_k = |A|^2 \left[
        N_a + \sum_{i\neq m} \exp\left[\mathrm{i}(\delta_{ik}-\delta_{mk})\right]
    \right].
\end{equation}
Taking the expectation value over the residual phases,
\begin{equation}
\label{eq:Ik_expect_appendix}
    \mathbb{E}[I_k]
    = |A|^2 \left[
        N_a + \sum_{i\neq m}
        \mathbb{E}\left(\exp\left[\mathrm{i}(\delta_{ik}-\delta_{mk})\right]\right)
    \right].
\end{equation}

Assume that the residual phases $\delta_{ik}$ are independent and identically
distributed draws from a symmetric distribution with zero mean. For any pair
$(i,m)$,
\begin{align}
    \mathbb{E}\left[\exp\left(\mathrm{i}(\delta_{ik}-\delta_{mk})\right)\right]
    &= \mathbb{E}\left[\exp(\mathrm{i}\delta_{ik})\right]\,
       \mathbb{E}\left[\exp(-\mathrm{i}\delta_{mk})\right] \nonumber\\
    &= \left|\mathbb{E}\left[\exp(\mathrm{i}\delta)\right]\right|^2,
\end{align}
where $\delta$ is a single residual phase draw. Define the complex phase
coherence
\begin{equation}
    \rho \equiv \left|\mathbb{E}\left[\exp(\mathrm{i}\delta)\right]\right|^2.
\end{equation}
There are $N_a(N_a-1)$ off-diagonal pairs $(i,m)$ in
Eq.~\ref{eq:Ik_expect_appendix}, so
\begin{equation}
    \mathbb{E}[I_k] = |A|^2 \left[ N_a + N_a(N_a-1)\,\rho \right].
\end{equation}

For a concrete example, let the residual phases be uniformly distributed on
$[-a,a]$,
\begin{equation}
    p(\delta) = \frac{1}{2a},\qquad -a \leq \delta \leq a.
\end{equation}
Then
\begin{align}
    \mathbb{E}\left[\exp(\mathrm{i}\delta)\right]
    &= \frac{1}{2a}\int_{-a}^{a} \exp(\mathrm{i}\delta)\,\mathrm{d}\delta \nonumber\\
    &= \frac{1}{2a}
       \left[\frac{\exp(\mathrm{i}\delta)}{\mathrm{i}}\right]_{-a}^{a}
     = \frac{\sin a}{a},
\end{align}
and the coherence is
\begin{equation}
    \rho = \left(\frac{\sin a}{a}\right)^2.
\end{equation}
The variance of a symmetric uniform distribution on $[-a,a]$ is
\begin{equation}
    \sigma_\delta^2 = \frac{a^2}{3},
\end{equation}
so $a = \sqrt{3}\,\sigma_\delta$. 
Then define
\begin{equation}
    f(\sigma_\delta) \equiv
    \frac{\sin(\sqrt{3}\,\sigma_\delta)}{\sqrt{3}\,\sigma_\delta},
\end{equation}
so that $\rho(\sigma_\delta) = f(\sigma_\delta)^2$. Substituting this into the
expression for $\mathbb{E}[I_k]$ gives
\begin{equation}
    \mathbb{E}[I_k]
    = |A|^2 \left[
        N_a + N_a(N_a-1)\,f(\sigma_\delta)^2
    \right].
\end{equation}
Normalising by the fully coherent gain $N_a^2 |A|^2$ leads to the dimensionless
array gain
\begin{equation}
    G(\sigma_\delta)
    \equiv \frac{\mathbb{E}[I_k]}{N_a^2 |A|^2}
    = \frac{1}{N_a}
      + \left(1-\frac{1}{N_a}\right) f(\sigma_\delta)^2,
\end{equation}
which is the curve plotted in Fig.~\ref{fig:coh_incoh_transition}. As
$\sigma_\delta\rightarrow 0$, $f(\sigma_\delta)\rightarrow 1$ and $G\rightarrow 1$,
corresponding to fully coherent phasing. As $\sigma_\delta$ grows and the
residual phases decorrelate, $f(\sigma_\delta)\rightarrow 0$ and
$G\rightarrow 1/N_a$, corresponding to the incoherent limit.

\section{Noise statistics of tied-array and incoherent beams}
\label{app:noise_stats}

In this appendix we derive the noise statistics used in Sect.~\ref{sec:flatfielding}.

\subsection{From complex Gaussian voltages to chi-squared powers}

Consider one station, one polarization, and one time–frequency sample in pure
noise. The complex voltage is
\begin{equation}
    V = X + \mathrm{i}Y,
\end{equation}
with $X,Y \sim \mathcal{N}(0,\sigma^2)$ independent. The instantaneous power in
this polarization is
\begin{equation}
    |V|^2 = X^2 + Y^2.
\end{equation}
Normalising by $\sigma^2$,
\begin{equation}
    \frac{|V|^2}{\sigma^2}
    = \left(\frac{X}{\sigma}\right)^2 + \left(\frac{Y}{\sigma}\right)^2,
\end{equation}
which is the sum of two squared standard normal variates, hence a $\chi^2_2$
variate. The total Stokes~$I$ power from both polarizations at a single station is
\begin{equation}
    I_{\mathrm{stat}} \propto |V_x|^2 + |V_y|^2
    = X^2 + Y^2 + U^2 + W^2,
\end{equation}
where $U,W$ are the real and imaginary parts of the second polarization. After
normalizing each component by $\sigma$, $I_{\mathrm{stat}}$ is proportional to a
$\chi^2_4$ variate.

For a TAB, complex voltages from $N_a$ stations are summed
with weights $\omega_i$,
\begin{equation}
    B_x = \sum_{i=1}^{N_a} \omega_i V_{x,i}, \qquad
    B_y = \sum_{i=1}^{N_a} \omega_i V_{y,i}.
\end{equation}
Because $B_x$ and $B_y$ are linear combinations of Gaussian voltages, they are
again complex Gaussian:
\begin{equation}
    B_x = X + \mathrm{i}Y,\quad
    B_y = U + \mathrm{i}W,
\end{equation}
with $X,Y,U,W$ independent zero-mean Gaussians of variance
\begin{equation}
    \sigma_B^2 = \sigma^2 \sum_{i=1}^{N_a} |\omega_i|^2.
\end{equation}
The tied-array Stokes~$I$ beam is
\begin{equation}
    I_{\mathrm{tab}} \propto |B_x|^2 + |B_y|^2
    = X^2 + Y^2 + U^2 + W^2.
\end{equation}
After normalizing by $\sigma_B^2$, this is the sum of four squared standard
normals, so
\begin{equation}
    I_{\mathrm{tab}} \;\text{follows a scaled}\; \chi^2_4 \;\text{distribution.}
\end{equation}
The coherent summation changes the scale (through $\sigma_B^2$) but not the number of degrees of freedom.

For an incoherent beam, a total power per station is first formed,
\begin{equation}
    P_i \propto |V_{x,i}|^2 + |V_{y,i}|^2,
\end{equation}
so each $P_i$ is a scaled $\chi^2_4$ variate. The incoherent sum of $N_a$ stations is
\begin{equation}
    I_{\mathrm{incoh}} \propto \sum_{i=1}^{N_a} P_i.
\end{equation}
The sum of $N_a$ independent $\chi^2_4$ variates is a $\chi^2$ variate with $4N_a$ degrees of freedom, so
\begin{equation}
    I_{\mathrm{incoh}} \;\text{follows a scaled}\; \chi^2_{4N_a} \;\text{distribution.}
\end{equation}
At the single-sample level, tied-array and incoherent beams therefore have different noise distributions: $\chi^2_4$ for TABs, $\chi^2_{4N_a}$ for
incoherent (up to calibration factors).

\subsection{Effect of averaging}

If $X$ follows a $\chi^2_k$ distribution, its mean and variance are
\begin{equation}
    \mu_X = k,\qquad
    \sigma_X^2 = 2k,
\end{equation}
and the fractional rms is
\begin{equation}
    \frac{\sigma_X}{\mu_X} = \sqrt{\frac{2}{k}}.
\end{equation}
If $M$ independent copies of $X$ are summed, the result follows $\chi^2_{kM}$,
and the average has mean $k$ and variance $2k/M$. The fractional rms of the
average is therefore
\begin{equation}
    \left(\frac{\sigma}{\mu}\right)_{\mathrm{avg}}
    = \sqrt{\frac{2}{kM}}.
\end{equation}

For the TAB, $k=4$, so after averaging $M$ independent
time–frequency samples
\begin{equation}
    \left(\frac{\sigma}{\mu}\right)_{\mathrm{tab}}
    = \sqrt{\frac{1}{2M}}.
\end{equation}
For the incoherent beam, $k=4N_a$, so
\begin{equation}
    \left(\frac{\sigma}{\mu}\right)_{\mathrm{incoh}}
    = \sqrt{\frac{1}{2N_a M}}.
\end{equation}
At fixed averaging scale $M$, an incoherent beam has smaller fractional noise fluctuations by a factor $\sqrt{N_a}$ and a distribution that is closer to Gaussian than the TAB. For realistic $N_a$ and $M$, both are well approximated as Gaussian in the central part of the distribution, but the TAB has more skew and heavier tails.

If a beam-averaged reference is then formed by averaging over $M_{\mathrm{beam}}-1$
beams that are in the incoherent regime, the effective degrees of freedom of the
reference increase by an additional factor $M_{\mathrm{beam}}-1$. Its fractional
rms is of order
\begin{equation}
    \left(\frac{\sigma_B}{\mu_B}\right)^2
    \approx \frac{1}{2N_a M (M_{\mathrm{beam}}-1)}.
\end{equation}
This is typically much smaller than the fractional rms of a single TAB, which is of order $1/(2M)$.

\subsection{beam flatfielding S/N with a noisy reference}

For an on-target sample in beam $k_0$ at fixed $(\nu,t)$, write
\begin{equation}
    I_{k_0} = \mu_{\mathrm{tab}} + S + n,\qquad
    \widehat{B}_{k_0} = \mu_B + \varepsilon,
\end{equation}
where $\mu_{\mathrm{tab}}$ is the mean noise level in the TAB, $S$ is
the signal contribution, $n$ is zero-mean TAB noise with variance
$\sigma_{\mathrm{tab}}^2$, and $\varepsilon$ is the zero-mean fluctuation of the
reference with variance $\sigma_B^2$. The flatfielded sample is
\begin{equation}
    \widetilde{I}_{k_0} = \frac{\mu_{\mathrm{tab}} + S + n}{\mu_B + \varepsilon}.
\end{equation}
Expanding to first order in $\varepsilon/\mu_B$,
\begin{equation}
    \widetilde{I}_{k_0}
    \approx \frac{\mu_{\mathrm{tab}} + S + n}{\mu_B}
    \left(1 - \frac{\varepsilon}{\mu_B}\right).
\end{equation}
Because the reference excludes beam $k_0$, $n$ and $\varepsilon$ can be treated as independent with zero covariance. The mean of $\widetilde{I}_{k_0}$ is
\begin{equation}
    \mathbb{E}[\widetilde{I}_{k_0}]
    \approx \frac{\mu_{\mathrm{tab}} + S}{\mu_B},
\end{equation}
so the signal amplitude is rescaled by $1/\mu_B$. The variance is
\begin{equation}
    \mathrm{Var}(\widetilde{I}_{k_0})
    \approx \frac{\sigma_{\mathrm{tab}}^2}{\mu_B^2}
    + \frac{(\mu_{\mathrm{tab}}+S)^2}{\mu_B^4}\,\sigma_B^2.
\end{equation}

Using the results above,
\begin{equation}
    \left(\frac{\sigma_{\mathrm{tab}}}{\mu_{\mathrm{tab}}}\right)^2
    \approx \frac{1}{2M},\qquad
    \left(\frac{\sigma_B}{\mu_B}\right)^2
    \approx \frac{1}{2N_a M (M_{\mathrm{beam}}-1)}.
\end{equation}
The ratio of the two noise contributions is then
\begin{equation}
    \frac{(\mu_{\mathrm{tab}}+S)^2 \sigma_B^2 / \mu_B^4}
         {\sigma_{\mathrm{tab}}^2 / \mu_B^2}
    \approx \frac{1}{N_a (M_{\mathrm{beam}}-1)}
    \left(\frac{\mu_{\mathrm{tab}}+S}{\mu_{\mathrm{tab}}}\right)^2.
\end{equation}
For realistic values $N_a \gtrsim 6$ and $M_{\mathrm{beam}}\gtrsim 40$ this factor is well below unity even when $S$ is comparable to $\mu_{\mathrm{tab}}$. The variance budget of the flatfielded beam is therefore dominated by the intrinsic TAB noise, not by the uncertainty in the reference.

Equivalently, the additional noise introduced by dividing by the empirical flatfield is suppressed by the large effective number of degrees of freedom in the reference. The main effect of beam flatfielding is to remove large-scale multiplicative structure, which reduces non-Gaussian wings in the noise distribution and improves the effective S/N for transient searches at fixed false-alarm rate.

\end{appendix}
\end{document}